\documentclass[11pt]{article}
\usepackage{booktabs}
\usepackage[preprint]{acl}

\usepackage{times}
\usepackage{latexsym}
\usepackage{tabularx}
\usepackage{makecell}
\usepackage{adjustbox}
\usepackage{float}
\usepackage[T1]{fontenc}
\usepackage{tabularx}
\usepackage{pifont}
\usepackage[utf8]{inputenc}
\usepackage{multirow}
\usepackage{microtype}

\usepackage{inconsolata}
\usepackage{pgfplots}
\pgfplotsset{compat=1.18}
\usepackage{graphicx}
\usepackage{array}
\usepackage{amsmath}
\usepackage{amssymb}       
\usepackage{pifont}        
\usepackage{xcolor, soul}
\usepackage{ulem}
\usepackage[utf8]{inputenc}
\usepackage{tcolorbox}
\usepackage{listings}
\usepackage{colortbl}
\tcbuselibrary{skins, breakable}
\newtcolorbox{promptbox}[1]{
    colback=white,
    colframe=cyan!70!black,
    fonttitle=\bfseries\large,
    title=#1,
    sharp corners=false,
    rounded corners,
    arc=3mm,
    boxrule=1pt,
    left=5pt,
    right=5pt,
    top=5pt,
    bottom=5pt,
    breakable, 
    enhanced
}
\usepackage{longtable}

\usepackage{tcolorbox}
\tcbuselibrary{skins, breakable}

\newcommand{\cmark}{\raisebox{0.2ex}{\textcolor[HTML]{009900}{\ding{51}}}}
\newcommand{\xmark}{\raisebox{0.2ex}{\textcolor[HTML]{FE0000}{\ding{55}}}}
\newcommand{\tmark}{\raisebox{0.15ex}{$\sim$}}
\newcolumntype{L}[1]{>{\raggedright\arraybackslash}p{#1}}
\newcolumntype{C}[1]{>{\centering\arraybackslash}p{#1}}
\newcolumntype{Y}{>{\raggedright\arraybackslash}X}

\title{InterveneBench: Benchmarking LLMs for Intervention Reasoning and Causal Study Design in Real Social Systems}

\author{
Shaojie Shi\thanks{Equal Contributions.}\textsuperscript{1,2} \quad
Zhengyu Shi\textsuperscript{*1} \quad
Lingran Zheng\textsuperscript{1,2} \quad
Xinyu Su\textsuperscript{1} \quad
Anna Xie\textsuperscript{1} \\
\textbf{Bohao Lv}\textsuperscript{1,2} \quad
\textbf{Rui Xu}\textsuperscript{1,2} \quad
\textbf{Zijian Chen}\textsuperscript{1} \quad
\textbf{Zhichao Chen}\textsuperscript{1} \quad
\textbf{Guolei Liu}\textsuperscript{1} \\
\textbf{Naifu Zhang}\textsuperscript{1} \quad
\textbf{Mingjian Dong}\textsuperscript{1} \quad
\textbf{Zhuo Quan}\textsuperscript{1} \quad
\textbf{Bohao Chen}\textsuperscript{1} \quad
\textbf{Teqi Hao}\textsuperscript{1} \\
\textbf{Yuan Qi}\textsuperscript{1,2} \quad
\textbf{Yinghui Xu}\thanks{Corresponding author.}\textsuperscript{1} \quad
\textbf{Libo Wu}\textsuperscript{$\dagger$,1,2} \\[2pt]
\textsuperscript{1} Fudan University \quad
\textsuperscript{2} Shanghai Innovation Institute \\
\texttt{sjshi24@m.fudan.edu.cn} \quad \texttt{zhengyushi96@gmail.com} \quad \texttt{lingran414@gmail.com}
}

\newcommand{\framework}{{STRIDES}}
\newcommand{\benchmark}{{InterveneBench}}

\begin{document}
\maketitle
\begin{abstract}
Causal inference in social science relies on end-to-end, intervention-centered research-design reasoning grounded in real-world policy interventions, but current benchmarks fail to evaluate this capability of large language models (LLMs). We present \textbf{\benchmark}, a benchmark designed to assess such reasoning in realistic social settings. Each instance in \benchmark\ is derived from an empirical social science study and requires models to reason about policy interventions and identification assumptions without access to predefined causal graphs or structural equations. \benchmark\ comprises 744 peer-reviewed studies across diverse policy domains. Experimental results show that state-of-the-art LLMs struggle under this setting. To address this limitation, we further propose a multi-agent framework, \textbf{\framework}. It achieves significant performance improvements over state-of-the-art reasoning models. Our code and data are available at \url{https://github.com/Sii-yuning/STRIDES}.

\end{abstract}

\section{Introduction}

Large language models (LLMs) have recently shown strong capabilities on causality-related tasks, such as causal discovery, causal graph generation, and counterfactual explanation \citep{kiciman2023causal}. However, most existing work frames causality as a closed and well-specified problem, assuming explicitly defined objectives, variables, and causal structures \citep{jin2023cladder,jin2024can,zhou2024causalbench}. In contrast, real-world social-science causal inference is open-ended and requires reasoning about interventions and their consequences under institutional and informational constraints \citep{imbens2015causal,athey2017state,ziems2024can}. Consequently, these challenges are not reflected in existing benchmarks, nor adequately addressed by current models \citep{liu-etal-2024-llms,acharya2025causcibench}.

\begin{figure}[!t]
    \centering
    \includegraphics[width=\columnwidth]{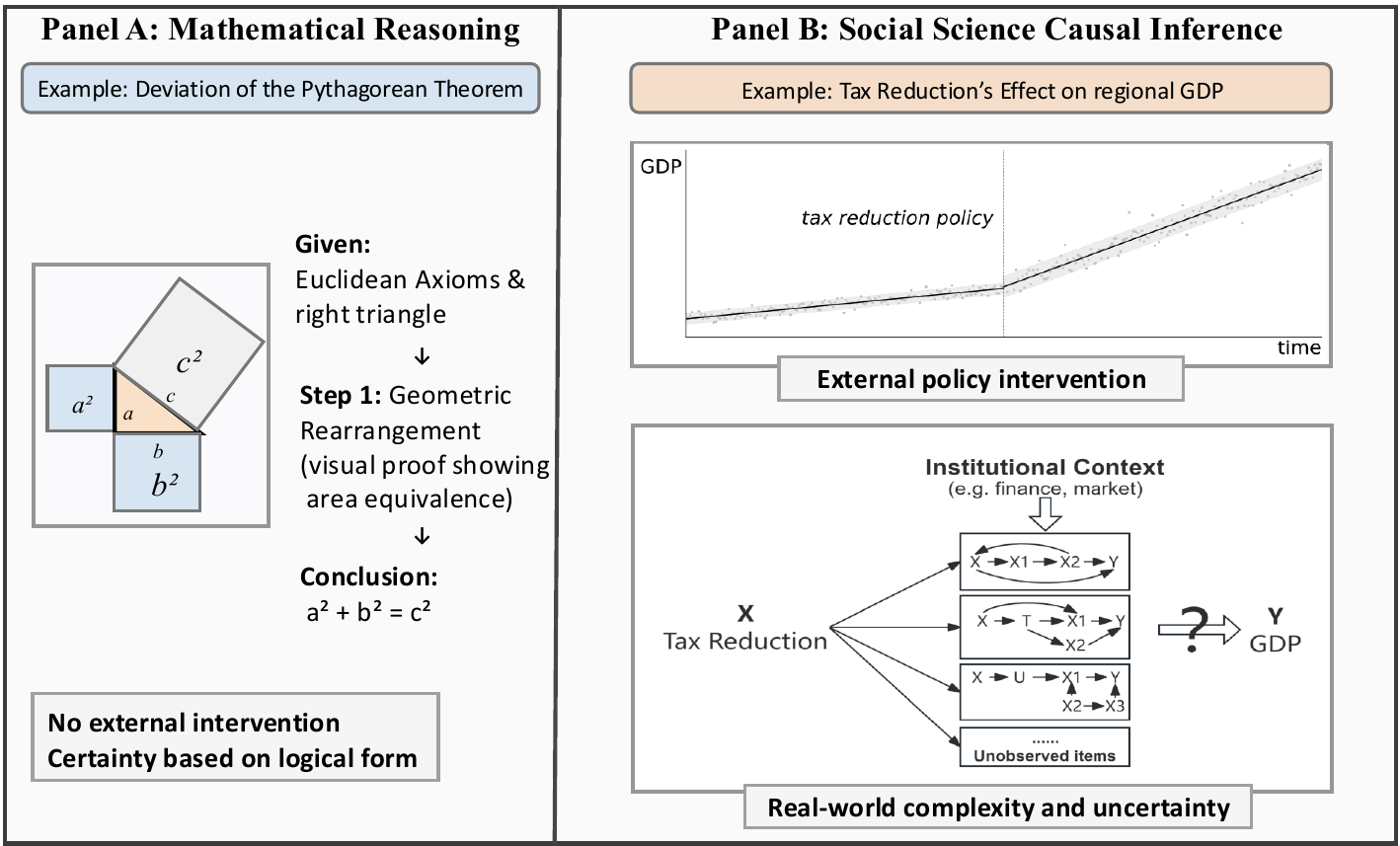}
    \vspace{-0.7cm}
    \caption{\small Comparison between closed-form mathematical reasoning (Panel A) and open-ended social-science causal inference (Panel B).}
    \label{fig:example}
\end{figure}

Figure~\ref{fig:example} highlights a key challenge of open-ended social causal inference: unlike closed-form reasoning, open-ended causal inference must operate under explicit interventions, absent predefined causal structures, and unobserved variables \citep{holland1986statistics,imbens2015causal}. Panel A describes a closed-form mathematical reasoning task in which objectives, primitives, and validity criteria are explicitly specified, allowing model performance to be evaluated by logical consistency or exact solution matching \citep{chen2021evaluatinglargelanguagemodels,cobbe2021training}. 
Panel B presents an example of social science causal inference, examining the effect of a tax reduction on regional GDP. In such tax reforms, various economic models can be used to infer causal effects, thus, the causal structure is not fixed a prior \citep{imbens2015causal,athey2017state}. Besides, in real-world policy settings, causal effects are shaped by diverse variables, some of which are hard to directly observe. 
As a result, causal reasoning in real-world policy contexts cannot be reduced to simple questions such as whether $X$ causes $Y$ \citep{Pearl2009}. Answering such questions requires an intervention-centered and structure-agnostic end-to-end causal study design, in which models reason about the intervention itself and make explicit assumptions under which causal conclusions can be justified. This requirement has direct implications for evaluation, as benchmarks designed for closed-form reasoning problems do not adequately assess whether models can perform the aforementioned kind of causal reasoning demanded by social-science research \citep{jin2023cladder,srivastava2023beyond,zhou2024causalbench,acharya2025causcibench}.

\begin{table*}[!t]
\centering
\scriptsize  
\setlength{\tabcolsep}{1.5pt}  
\renewcommand{\arraystretch}{1.0}  

\begin{tabular*}{\textwidth}{@{\extracolsep{\fill}}@{}
>{\raggedright\arraybackslash}p{4cm}
>{\centering\arraybackslash}p{1.8cm}
>{\centering\arraybackslash}p{1.5cm}
>{\centering\arraybackslash}p{1.8cm}
>{\centering\arraybackslash}p{1.8cm}
>{\centering\arraybackslash}p{3.3cm}
>{\centering\arraybackslash}p{1.0cm}
@{}}
\toprule
\textbf{Benchmark} &
\parbox[c]{1.8cm}{\centering\textbf{End-to-End}\\\textbf{Causal}\\\textbf{Study Design}} &
\parbox[c]{1.5cm}{\centering\textbf{Real-World}\\\textbf{Reasoning}} &
\parbox[c]{1.8cm}{\centering\textbf{Interventional}\\\textbf{Reasoning}} &
\parbox[c]{1.8cm}{\centering\textbf{Structure-}\\\textbf{Agnostic}\\\textbf{Reasoning}} &
\textbf{Source} &
\textbf{Size} \\
\midrule

GSM8K \citep{cobbe2021training} & \cmark & \xmark & \xmark & \xmark &
Human-written & 1K \\

HumanEval \citep{chen2021evaluatinglargelanguagemodels} & \cmark & \xmark & \xmark & \xmark &
Human-written & 164 \\

MMLU \citep{hendrycks2021measuring} & \cmark & \tmark & \xmark & \xmark &
Professional exams & $\sim$14K \\
\midrule  

DiscoveryBench \citep{majumder2025discoverybench} & \xmark & \tmark & \xmark & \xmark &
Papers + Synthetic scenarios & $\sim$420 \\

QRData \citep{liu-etal-2024-llms} & \xmark & \cmark & \tmark & \xmark &
Textbooks + Web + Papers & $\sim$701 \\


CLADDER \citep{jin2023cladder} & \cmark & \cmark & \tmark & \xmark &
Books + Papers & 10K \\

CauSciBench \citep{acharya2025causcibench} & \cmark & \tmark & \tmark & \xmark &
Papers + Refined Data & 305 \\

\midrule
\textbf{InterveneBench} & \cmark & \cmark & \cmark & \cmark &
\textbf{Papers from SCI / SSCI} & \textbf{744} \\
\bottomrule
\end{tabular*}

\caption{\small Comparison of benchmarks.\textbf{End-to-End Causal Study Design} evaluates causal reasoning as a complete social science inquiry, from problem formulation and intervention specification to causal identification and outcome interpretation. \textbf{Real-World Reasoning} reflects grounding in real policy interventions. \textbf{Interventional Reasoning} presents reasoning about interventions. \textbf{Structure-Agnostic Reasoning} denotes that no explicit causal graphs or structural equations are provided, requiring reasoning under open-ended institutional contexts.}
\label{tab:final-comparison}
\end{table*}

To bridge this gap, we present \benchmark, a benchmark for evaluating whether model reasoning can support an intervention-centered and structure-agnostic end-to-end causal study design under real-world policy interventions. As summarized in Table~\ref{tab:final-comparison}, \benchmark\ is the first benchmark to jointly evaluate four dimensions that are critical for social-science causal inference but largely absent from existing benchmarks. Specifically, it can evaluate models to (i) conduct end-to-end causal study design rather than solve isolated subtasks, (ii) reason about interventions grounded in empirical policy studies and institutional contexts, (iii) perform interventional reasoning about policy-induced changes rather than correlations, and (iv) operate in a structure-agnostic setting without access to explicit causal graphs or structural equations.

Constructing such a benchmark requires grounding causal reasoning in real-world settings rather than synthetic tasks~\citep{wang2024scibench}.
We therefore draw on empirical social science studies and develop a multi-agent pipeline to extract and reconstruct research-grade causal study designs from high-quality papers~\citep{hong2023metagpt,wu2024autogen}.
A Paper Interpreter agent parses the metadata like study context and interventions. A Causal Designer agent incorporates the causal assumptions and estimates, as well as the corresponding identification strategy described in the paper. And a Critic agent cross-checks all generated content against the source text \citep{madaan2023self,shinn2023reflexion}. All instances are further expert-verified to ensure that \benchmark\ reflects research-grade causal reasoning.

We further evaluate a set of state-of-the-art (SOTA) LLMs on \benchmark. Results show that current models still fall short of social causal study design. Even the strongest model, GPT-5.1, achieves an aggregate score of 0.578, and only 49.3\% accuracy on model selection. These results highlight persistent weaknesses in current LLMs’ capacity to support end-to-end real-world study design and interventional reasoning.

Motivated by this gap, we further introduce \framework\ (\textbf{S}ocial \textbf{T}heory-guided \textbf{R}esearch for \textbf{I}ntervention \textbf{D}esign, \textbf{E}stimation, and \textbf{S}crutiny), a multi-agent system tailored for social science causal inference. STRIDES mirrors expert collaboration by decomposing the workflow into specialized roles for design, checking, and refinement.

Our contributions are summarized as follows:

-- We introduce \benchmark, the first benchmark to evaluate end-to-end causal study design reasoning under real-world policy interventions. 

-- We empirically demonstrate that SOTA LLMs struggle on \benchmark, despite strong performance on existing causal and reasoning benchmarks. We further propose \framework, a social theory-guided multi-agent framework that simulates expert collaboration in social-science research. \framework\ significantly improves LLM performance on \benchmark.

-- We conduct experiments on \benchmark, integrating our proposed \framework\ with current LLMs. The results show that using \framework\ consistently improves performance over corresponding baselines by up to 25.1\% in Final Score.



\begin{figure*}[t]
    \centering
    \includegraphics[width=\textwidth]{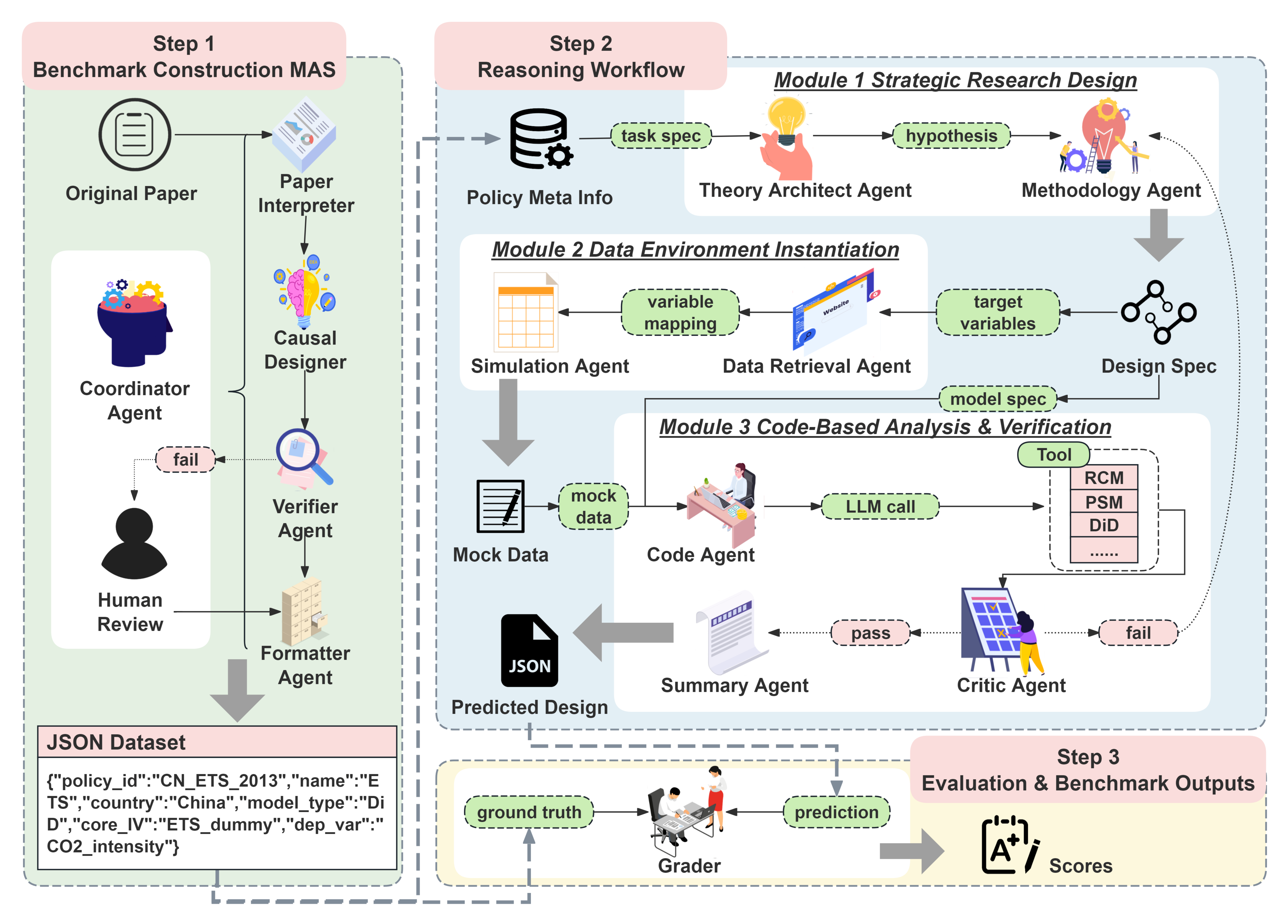}
    \vspace{-1cm}
    \caption{\small Overview of our proposed system. The system has three stages. (1)~Benchmark construction uses a Human-in-the-Loop MAS: a coordinator schedules a Paper Interpreter and Causal Designer to produce a draft causal design, a Verifier checks and routes low-quality designs for human review, and a Formatter converts approved designs into a standardized JSON schema as ground-truth designs. (2)~\framework\ mirrors expert research workflows through three sequential modules.~(i)~The Strategic Research Design module maps unstructured metadata to statistical models using two specialized agents.~(ii)~The Data Environment Instantiation module employs a data retrieval agent to derive measurable indicators from model specifications and a simulation agent to generate mock data.~(iii)~The Code-Based Analysis and Verification module uses a code agent to produce executable statistical code and a critic agent to provide iterative feedback for design refinement, followed by a summary agent that standardizes qualified predicted designs.~(3)~Evaluation is performed by a separate grader module (not part of the generation pipeline), which compares model predictions against expert-verified ground truth and outputs evaluation scores.}
    \label{fig:framework_wide}
\end{figure*}

\section{Related Work}

\subsection{Causal Reasoning Benchmarks for LLMs}

The remarkable zero-shot reasoning and world-knowledge synthesis capabilities of LLMs have demonstrated significant promise in causal inference~\citep{kiciman2023causal}. However, existing benchmarks predominantly focus on closed-form, well-specified problems such as Mathematics~\citep{cobbe2021training} or Code Generation~\citep{chen2021evaluatinglargelanguagemodels}, which remain disconnected from real-world open-ended scientific research (Table~\ref{tab:final-comparison}).

Although recent works have begun to explore interventional and counterfactual reasoning, current benchmarks still fall short of modeling the full scientific workflow. QRData~\citep{liu-etal-2024-llms} focuses on data-driven statistical estimation without formulating domain-specific scientific inquiries. DiscoveryBench~\citep{majumder2025discoverybench} and ResearchBench~\citep{DBLP:journals/corr/abs-2503-21248} facilitate hypothesis generation yet remain confined to the initial discovery phase, primarily evaluating isolated components of research workflows rather than end-to-end scientific reasoning.

While some researchers consider a full workflow, they usually adopt predefined structures and datasets, limiting models' flexibility. CLADDER~\citep{jin2023cladder} introduces a comprehensive reasoning chain including interventional effects, but it is strictly grounded in predefined graphical models or structural causal models (SCMs). Even the most advanced end-to-end framework, CauSciBench~\citep{acharya2025causcibench}, remains dependent on data given in advance and focuses on calculation, notably lacking preliminary assumption design and critical scientific interpretation.

Crucially, causal inference in social science necessitates a structure-agnostic research design, which should autonomously synthesize the end-to-end pipeline---including theoretical framework design, data retrieval, statistical calculation, and interpretation---while being grounded solely in background contexts and real-world policy interventions. Current benchmarks rely on predefined equations or datasets, rendering them incapable of modeling the complete, iterative workflow of social science causal inference.

To address this gap, we construct a benchmark from published studies. Our framework incorporates an end-to-end logical chain, a structure-free analytical pipeline, and statistically verifiable real-world results, thereby providing a more challenging and credible evaluation for LLM-based causal reasoning.

\subsection{LLM-based Multi-Agent Systems}

LLM-based Multi-Agent Systems (MAS) have emerged as a powerful paradigm for enhancing complex reasoning by leveraging distributed inference, hierarchical planning, and collaborative mechanisms that mirror human expert workflows. Owing to its superior complex reasoning abilities, LLM-based MAS demonstrates remarkable performance in long-context understanding and fine-grained metadata extraction.

CoA~\citep{zhang2024chain} employs a sequential multi-agent collaboration strategy where agents process segmented long text to extract informational snippets, which are then synthesized by a ``manager'' agent into a coherent response. DocLens~\citep{zhu2025doclens} utilizes a tool-augmented multi-agent architecture to localize and extract fine-grained visual and textual evidence from long visual documents (e.g., multi-page reports with charts and tables). KARMA~\citep{lu2025karma} employs collaborative agents to parse unstructured literature and extract entities for domain-specific Knowledge Graph construction. The success of these multi-agent frameworks shows that LLM-based MAS can perform long-context reasoning and precise metadata extraction from academic literature by leveraging specialized agent roles.

However, existing studies have not explored how MAS can support LLMs in reasoning about open-ended causal study design. Our work bridges these two lines of research by constructing an intervention-centric, structure-agnostic end-to-end causal study benchmark, and further augmenting the social causal reasoning capabilities of LLMs through our proposed MAS framework \framework.

\section{Preliminaries} 
\paragraph{Theoretical Grounding.} Intervention-based notions of causality are widely adopted in social and economic systems. Unlike approaches that focus on recovering the graph structures of static Structural Causal Models (SCMs), causal inference in social science is commonly formulated through comparisons of potential outcomes under different treatment statuses.

Importantly, causal validity does not require strict "ceteris paribus" (all other things being equal). Instead, each identification strategy relies on specific, testable assumptions, e.g., parallel trends, balance test, or local randomization at a cutoff~\citep{ding2025can,chen2025laissez,hammarlund2025impact}, that make remaining group differences unlikely to bias causal estimates.


\paragraph{Problem Setting.} Given a Policy Intervention $\mathcal{T}$ (e.g. tax reform, medical insurance, student loan, etc.), the observed outcome variable $Y$ reflects as aggregate effect of the system under the intervention. In our setting, $Y$ emerges from multiple LLM-driven expert agents as they reason about $\mathcal{T}$.
Consequently, our task aims to establish a rigorous identification strategy to effectively estimate the average treatment effect $\tau$ of policy interventions, defined in terms of potential outcomes:

{\small
\begin{equation}
    \tau = E[Y(\mathcal{T}=1)- Y(\mathcal{T}=0)].
\end{equation}
}
\paragraph{Task Definition.}
Based on the problem setting, we formalize intervention-based multi-agent causal reasoning task as a generation and verification problem, mapping unstructured metadata to structured research designs. The input can be represented as a set containing policy meta-information:

{\small
\begin{equation}
    \mathcal{M} = \{c_{inst}, t_{policy}, o_{aim}, \dots\},
\end{equation}
}
where $c_{inst}$ describes the institutional context, $t_{policy}$ details the specific intervention, and $o_{aim}$ outlines the research objective.

The output is represented as an optimal research design scheme $\mathcal{D}^*$, defined as a quadruple:

{\small
\begin{equation}
    \mathcal{D} = \langle S_{id}, V_{spec}, \mathcal{A}_{ssump}, \pi_{valid} \rangle.
\end{equation}
}

Here, $S_{id}$ denotes the identification strategy (e.g., difference-in-differences (DiD), regression discontinuity (RD), instrumental variables (IV)); $V_{spec}$ specifies the Variable Specifications (including the core independent variable, outcome variable, and set of covariates), $\mathcal{A}_{ssump}$ represents the Critical Assumptions required for the identification strategy to hold (e.g., parallel trends, exclusion restrictions) and $\pi_{valid}$ denotes the Statistical Testing Logic used to validate the design.

Our goal is to construct a multi-agent system $\Phi$ such that the generated $\mathcal{D}$ maximally approximates the Ground Truth design $\mathcal{D}_{gt}$, which has been validated in authentic social science research:

{\small
\begin{equation}
    \Phi^* = \mathop{\arg\max}_{\Phi} P(\mathcal{D} \approx \mathcal{D}_{gt} \mid \mathcal{M}, \mathcal{K}_{domain}),
\end{equation}
}
where $\mathcal{K}_{domain}$ represents the externally introduced domain knowledge base.

\section{Methodology}
To systematically evaluate the performance of LLMs on intervention-centered, structure-agnostic causal reasoning in social science, we first introduce a benchmark dataset, named \textbf{\benchmark}. This dataset is designed to evaluate reasoning as end-to-end causal study design under real-world policy interventions. Using \benchmark, we conduct a comprehensive evaluation of existing LLMs to assess their reasoning abilities in this setting. We further propose a generic framework, named \textbf{\framework}, which aims to enhance LLMs’ performance on intervention-based causal reasoning tasks in social science.

%

\subsection{Benchmark Construction}
\label{sec:benchmark_con}
To facilitate the task, we aim to construct a high-fidelity benchmark capable of effectively assessing causal reasoning capabilities. To mitigate the cost and scalability limitations of manual annotation, we propose a semi-automated construction framework incorporating Human-in-the-Loop (HITL) mechanism. This framework uses a collaborative multi-agent system (MAS) comprising five agents (the Coordinator Agent, Paper Interpreter Agent, Causal Design Agent, Verifier Agent and Formatter Agent) to transform unstructured text into structured causal designs, followed by expert verification to ensure data quality. Step 1 in Figure~\ref{fig:framework_wide} presents this benchmark construction MAS.

\paragraph{Coordinator Agent.} This agent serves as the central control node and manages the state machine of the overall workflow. It schedules sub-agents via a predefined directed acyclic graph~(DAG)~and 
handles error recovery in response to anomalies such as formatting errors or API timeouts. In addition, it maintains a global context $\mathcal{C}_{global}$ to ensure consistent information propagation among Interpreter, Designer, and Verifier agents.

\paragraph{Paper Interpreter Agent.} This agent uses \texttt{pypdf} to parse raw PDF documents and perform the preliminary structured extraction from unstructured text. The output, denoted as $\mathcal{M}_{raw}$, includes the paper's raw content, covering policy background, implementation timelines, and data sources.

\paragraph{Causal Designer Agent.} This agent performs deep reverse engineering over $\mathcal{M}_{raw}$ to map natural language descriptions to core elements of social science causal inference. Specifically, it (1)~ extracts explicit social model (e.g., difference-in-differences), and (2)~ reconstructs the author's implicit design logic (e.g., control group selection criteria or variable construction methods). The output, denoted as $\mathcal{D}_{raw}$, represents a preliminary ground-truth causal design.

\paragraph{Verifier Agent.} This agent performs a self-consistency check to assess whether the generated causal design $\mathcal{D}_{raw}$ is logically entailed by the original textual evidence $E_{text}$, and outputs a corresponding confidence score. Formally, we define:

{\small
\begin{equation}
    Score_{ver} = A_{ver}(\mathcal{D}_{raw}, E_{text}),
\end{equation}
}
where $A_{ver}$ outputs a scalar $s \in [0,1]$. A data entry is deemed machine-verified only if $Score_{ver}$ exceeds a preset threshold $\lambda$ (set to 0.9 in this paper); otherwise, the sample is flagged and routed to a \textbf{Human Review Queue}, where domain experts verify and refine it, and then re-integrated into the dataset.

\paragraph{Formatter Agent.} This agent standardizes and cleans the data by mapping the approved $\mathcal{D}_{raw}$ to a predefined JSON schema $\mathcal{D}_{gt}$. It unifies units and variable naming conventions and handles missing values, ensuring that the final benchmark dataset is both machine-readable and consistent.

To ensure the benchmark is robust and meaningful to researchers in the target social science domains, we incorporate extensive human expert validation throughout the construction process. Four domain experts with graduate-level training in economics and public policy independently review each instance, verifying causal design completeness, variable operationalization, and identification strategy appropriateness. We further validate benchmark robustness through cross-model extraction consistency checks (§\ref{app_sec:extraction_consistency}), LLM-judge stability analysis under semantic paraphrasing (Appendix~\ref{app_case_con}), and re-weighting sensitivity tests (Appendix~\ref{app_sec:reweighting}), collectively demonstrating that the benchmark yields stable, expert-aligned evaluations.

\subsection{The STRIDES Framework}
To address the limitations of SOTA LLMs in social science causal inference, we propose \framework, a collaborative multi-agent framework that mimics the research workflow of human experts (see Step 2 in Figure~\ref{fig:framework_wide}). This framework decomposes complex strategic planning tasks into three sequential modules featuring internal iterative feedback: Strategic Research Design, Data Environment Instantiation, and Code-Based Analysis and Verification, implemented by seven specialized agents.

\paragraph{Strategic Research Design Module.}
This module addresses the mapping from unstructured metadata to statistical models. To prevent the generation of statistically feasible but theoretically flawed results, we introduce a mechanism for Theory-Method Decoupling.

\textit{Theory Architect Agent ($A_{theo}$).} This agent serves as the logical starting point of the system and is responsible for constructing a qualitative causal narrative. It takes metadata $\mathcal{M}$ as input and utilizes an embedded knowledge base from economics and sociology (e.g., supply and demand theory, institutional change theory) to formulate theoretical hypotheses $\mathcal{H}_{hyp}$. This step does \emph{not} involve specific statistical modeling; instead, it focuses on the logical consistency of the causal chain:

{\small
    \begin{equation}
        \mathcal{H}_{hyp} = A_{theo}(\mathcal{M}).
    \end{equation}
}

\textit{Methodology Agent ($A_{meth}$).} This agent aims to translate qualitative hypotheses $\mathcal{H}_{hyp}$ into mathematical identification strategies. It determines the specific model specification $S_{id}$, clearly defining the treatment variable, outcome variable, and control strategies (such as the selection of fixed effects) to eliminate confounders:

{\small
    \begin{equation}
        S_{id} = A_{meth}(\mathcal{H}_{hyp}, \mathcal{M}).
    \end{equation}
}
\paragraph{Data Environment Instantiation Module.}
This module enables the model to reason and think within concrete data constraints.

\textit{Data Retrieval Agent ($A_{retr}$).} This agent maps abstract concepts in $S_{id}$ to measurable real-world indicators $V_{core}$ and verifies data availability via external knowledge bases to avoid invalid designs due to data access issues.

\textit{Simulation Agent ($A_{sim}$).} This agent simulates how researchers process data when designing a causal study. It helps the LLM understand how key variables are numerically related by creating a controlled, virtual data setting in which plausible effects and noise are explicitly introduced.
Specifically, it generates a synthetic logical dataset (\emph{Mock Data}, $D_{mock}$) over the statistical characteristics of $V_{core}$ (distribution, trends, and dimensionality) to provide a runtime environment for subsequent verification. Formally,

{\small
    \begin{equation}
        D_{mock} \sim P_{sim}(D \mid V_{core}).
    \end{equation}
}
\paragraph{Code-Based Adversarial Verification Module.}
This module consists primarily of two agents functioning in an adversarial loop.

\textit{Code Agent ($A_{code}$).} This agent translates the design specification $S_{id}$ into executable statistical code $\pi_{code}$ (e.g., in Python using \texttt{statsmodels} or R) and executes it on $D_{mock}$ to produce results $\mathcal{R}_{exec}$, which includes regression coefficients, standard errors, significance levels, and potential runtime errors. Common social models, including DiD, PSM, and IV, are encapsulated as callable tools for the LLM. Formally:

{\small
    \begin{equation}
        \mathcal{R}_{exec} = \text{Execute}(\pi_{code}, D_{mock}).
    \end{equation}
}

\textit{Critic Agent ($A_{critic}$).} This agent performs a consistency review between $\mathcal{R}_{exec}$ and $\mathcal{H}_{hyp}$. It focuses on two types of issues:
~(i)~\emph{Statistical Pitfalls,} including Multicollinearity, non-convergence, or failure of parallel trends tests, and~(ii)~\emph{Logical Deviations,} where execution outcomes (e.g., coefficient signs) contradict theoretical assumptions.
The Critic Agent outputs a feedback signal $f_{feedback}$. If a critical error is detected, the information is propagated back to $A_{meth}$ or $A_{code}$ for iterative refinement until the design passes review or reaches the maximum iteration limit:

{\small
    \begin{equation}
        \text {If}\ \text{Reject}(A_{critic}(\mathcal{R}_{exec})) \rightarrow \text{Refine}(S_{id} \text{ or } \pi_{code})
    \end{equation}
}
\textit{Summary Agent ($A_{sum}$).} This agent converts  $\mathcal{R}_{\text{exec}}$ into a predicted causal design represented in a predefined JSON schema $\mathcal{D}$. 

The Grader computes scores by comparing predicted causal designs against ground-truth to form a closed-loop mechanism that enhances \framework’s performance on social-science causal inference. For case study of the consistency of the Grader, please refer to Appendix \ref{app_case_con}.

\section{Experiments}

\subsection{Benchmark}
\label{app:dataset}
\benchmark\ consists of 744 empirical social science studies, each corresponding to a policy intervention and its associated causal study design. For each instance, the benchmark specifies the policy intervention, relevant background environment, and the identification strategy employed in the original study, forming a complete unit for evaluating end-to-end causal study design reasoning.

As shown in Table~\ref{tab:data_stats}, \benchmark\ covers five major causal inference paradigms, including DiD, IV, RD, SCM (Synthetic Control Method), and PSM (Propensity Score Matching), and nine policy domains spanning environment, economics, health, education, etc. This diversity enables systematic evaluation of model performance across methodological choices and substantive policy contexts.

To use \benchmark\ for evaluation, models are tasked with reasoning about the causal effect of a given policy intervention and proposing a study design under the provided context. We adopt a hierarchical rubric comprising three parts and spanning three dimensions, with a total of 45 raw points normalized to $[0, 1]$.

The rubric allows partial credit for partially aligned components rather than treating any single mistake as an all-or-nothing failure. Full scoring details are provided in Appendix~\ref{app_sec:metric}. Because many interventions have been implemented in the real world, evaluation is susceptible to knowledge-cutoff effects, whereby LLMs may rely on memorized pre-training content. To mitigate this risk, we build a time-restricted test set consisting only of papers published in 2025, yielding 74 held-out instances, while the remaining 670 papers form the Legacy Set. We also conduct experiments on the Legacy Set, whose results are reported in Appendix ~\ref{app:result_lag}.


\subsection{Experimental Setup}

\paragraph{Dataset \& Baselines.} 
We use \benchmark\ to evaluate the causal reasoning capabilities of a diverse set of SOTA proprietary and open-source LLMs. Specifically, the evaluated models include GPT-5.1 \citep{openai_gpt51}, Claude Sonnet 4 \citep{anthropic_claude4},
and Claude 3.7 Sonnet \citep{anthropic_claude37}, Gemini-2.5-Pro \citep{comanici2025gemini}, Grok-4 \citep{xai2025grok4modelcard}, Kimi-K2 \citep{team2025kimi}, Gemini-3-Flash \citep{googlecloud2025gemini3flash}, DeepSeek-v3.2 \citep{liu2025deepseek}, GLM-4.6 \citep{zhipu2026glm46}, GPT-OSS-120B \citep{agarwal2025gpt} and Qwen3-235B-A22B \citep{yang2025qwen3}. Beyond vanilla inference, we evaluate three additional baselines---Few-shot (3-shot), CoT + Self-Refine, and Best-of-6---detailed in Appendix~\ref{app_sec:stronger_baselines}. Prompts are in Appendix~\ref{sec:prompts}.

\paragraph{Implementation.}
\framework\ is implemented using the LangGraph framework \citep{langgraph_2025_ga}, which provides a DAG structure for coordinating state transitions and message passing among agents. Regarding hyperparameter settings, for agents responsible for logical reasoning, we set the temperature to 0 to ensure deterministic outputs. For other agents, we set the temperature to 0.7 to enhance the diversity and robustness. The raw score is normalized to $[0,1]$. 



\begin{table}[!t]
\centering
\small
\setlength{\tabcolsep}{4pt}
\renewcommand{\arraystretch}{1.08}
\resizebox{\columnwidth}{!}{
\begin{tabular}{l r r}
\toprule
\multicolumn{3}{l}{\textbf{(A) causal Inference Model Distribution}}\\
\midrule
Model type & Legacy Set & Test Set \\
\midrule
Difference-in-Differences (DiD) & 241 & 42 \\
Instrumental Variables (IV) & 203 & 15 \\
Regression Discontinuity (RD) & 113 & 7 \\
Synthetic Control Method (SCM) & 93 & 7 \\
Propensity Score Matching (PSM) & 20 & 3 \\
\midrule
\textbf{Total} & \textbf{670} & \textbf{74} \\
\midrule\midrule
\multicolumn{3}{l}{\textbf{(B) Policy Domain Distribution}}\\
\midrule 
Policy domain & Legacy Set & Test Set \\
\midrule
Environmental \& Sustainable Development & 16 & 2 \\
Housing \& Urban Development & 79 & 5 \\
Social Security \& Welfare & 47 & 13 \\
Science, Technology, \& Innovation & 120 & 14 \\
Public Health & 91 & 7 \\
Education \& Training & 46 & 3 \\
Labor \& Employment & 53 & 10 \\
Macroeconomic & 208 & 17 \\
Income Distribution & 10 & 3 \\
\midrule
\textbf{Total} & \textbf{670} & \textbf{74} \\
\bottomrule
\end{tabular}
}
\vspace{-0.2cm}
\caption{\small \benchmark\ statistics by causal inference model type and policy domain. We construct a time-restricted test split using 2025 papers to mitigate knowledge-cutoff contamination risks.}
\label{tab:data_stats}
\end{table}

\begin{table*}[!t]
\centering
\small
\setlength{\tabcolsep}{5pt}
\renewcommand{\arraystretch}{1.1}
\resizebox{\textwidth}{!}{
\begin{tabular}{l|cccccccc|c}
\toprule
\textbf{Model} & \textbf{Final Score} & \textbf{Model Type} & \textbf{Core IV} & \textbf{Group Def} & \textbf{Controls} & \textbf{Dep Var} & \textbf{Reasoning} & \textbf{Explanation} & \textbf{Improve} \\
\midrule

\textbf{STRIDES(GPT-5.1)} & \textbf{{\color{red}0.665}} & \textbf{0.515} & \textbf{0.691} & \textbf{0.544} & \textbf{{\color{red}0.842}} & \textbf{{\color{red}0.974}} & \textbf{{\color{red}0.695}} & \textbf{{\color{red}0.650}}  & \multirow{2}{*}{+15.1\%} \\
\quad \textit{w/o MAS} & \underline{0.578} & 0.493 & \underline{0.656} & 0.535 & \underline{0.652} & \underline{0.658} & \underline{0.530} & \underline{0.517} \\
\midrule

\textbf{STRIDES(Claude-3.7-Sonnet)} & \textbf{0.653} & \textbf{{\color{red}0.650}} & \textbf{0.728} & \textbf{0.601} & \textbf{0.570} & \textbf{0.892} & \textbf{0.620} & \textbf{0.347} & \multirow{2}{*}{+20.0\%}\\
\quad \textit{w/o MAS} & 0.544 & 0.519 & 0.578 & 0.535 & 0.570 & 0.656 & 0.510 & 0.340 \\
\midrule

\textbf{STRIDES(Claude-Sonnet-4)} & \textbf{0.652}& \textbf{0.634} & \textbf{0.734} & \textbf{{\color{red}0.652}} & \textbf{0.550} & \textbf{0.816} & \textbf{0.615} & \textbf{0.367} & \multirow{2}{*}{+21.0\%}\\
\quad \textit{w/o MAS} & 0.539 & \underline{0.541} & 0.573 & \underline{0.541} & 0.488 & 0.656 & \underline{0.530} & 0.307 \\
\midrule

\textbf{STRIDES(GLM-4.6)} & \textbf{0.642} & \textbf{0.606} & \textbf{0.709} & \textbf{0.646} & 0.464 & \textbf{0.882} & \textbf{0.600} & \textbf{0.453} & \multirow{2}{*}{+20.9\%} \\
\quad \textit{w/o MAS} & 0.531 & 0.535 & 0.566 & 0.533 & \textbf{0.480} & 0.644 & 0.520 & 0.303 \\
\midrule

\textbf{STRIDES(Gemini-2.5-Pro)} & \textbf{0.621} & \textbf{0.646} & \textbf{{\color{red}0.762}} & \textbf{0.578} & 0.500 & \textbf{0.740} & \textbf{0.600} & 0.233 & \multirow{2}{*}{+24.9\%}\\
\quad \textit{w/o MAS} & 0.497 & 0.477 & 0.509 & 0.477 & \textbf{0.508} & 0.594 & 0.475 & \textbf{0.427} \\
\midrule

\textbf{STRIDES(Gemini-3-Flash)} & \textbf{0.583} & \textbf{0.525} & \textbf{0.638} & \textbf{0.586} & \textbf{0.460} & \textbf{0.750} & \textbf{0.530} & \textbf{0.547} & \multirow{2}{*}{+25.1\%} \\
\quad \textit{w/o MAS} & 0.466 & 0.445 & 0.477 & 0.477 & 0.434 & 0.552 & 0.445 & 0.393 \\
\midrule

\textbf{STRIDES(Grok-4)} & \textbf{0.580} & \textbf{0.550} & \textbf{0.657} & \textbf{0.591} & 0.436 & \textbf{0.770} & \textbf{0.550} & 0.327 & \multirow{2}{*}{+25.0\%}\\
\quad \textit{w/o MAS} & 0.464 & 0.424 & 0.514 & 0.445 & \textbf{0.462} & 0.572 & 0.420 & \textbf{0.340} \\
\midrule

\textbf{STRIDES(GPT-4.1)} & \textbf{0.570} & \textbf{0.541} & \textbf{0.646} & \textbf{0.580} & 0.430 & \textbf{0.756} & \textbf{0.540} & \textbf{0.320} & \multirow{2}{*}{+25.0\%} \\
\quad \textit{w/o MAS} & 0.456 & 0.445 & 0.498 & 0.430 & \textbf{0.458} & 0.572 & 0.445 & 0.243 \\
\midrule

\textbf{STRIDES(Qwen3-235B-A22B)} & \textbf{0.569} & \textbf{0.538} & \textbf{0.638} & \textbf{0.544} & \textbf{0.534} & \textbf{0.712} & \textbf{0.525} & \textbf{0.373} & \multirow{2}{*}{+18.0\%}\\
\quad \textit{w/o MAS} & 0.482 & 0.456 & 0.541 & 0.461 & 0.452 & 0.604 & 0.445 & 0.317 \\
\midrule

\textbf{STRIDES(GPT-OSS-120B)} & \textbf{0.553} & \textbf{0.553} & \textbf{0.611} & \textbf{0.512} & 0.436 & \textbf{0.726} & \textbf{0.550} & \textbf{0.413} & \multirow{2}{*}{+24.8\%}\\
\quad \textit{w/o MAS} & 0.443 & 0.435 & 0.440 & 0.403 & \textbf{0.466} & 0.572 & 0.415 & 0.377 \\
\midrule

\textbf{STRIDES(Kimi-K2)} & \textbf{0.519} & \textbf{0.523} & \textbf{0.555} & \textbf{0.484} & 0.424 & \textbf{0.674} & \textbf{0.480} & \textbf{0.420} & \multirow{2}{*}{+25.1\%}\\
\quad \textit{w/o MAS} & 0.415 & 0.361 & 0.445 & 0.414 & \textbf{0.438} & 0.508 & 0.360 & 0.340 \\
\midrule

\textbf{STRIDES(DeepSeek-v3.2)} & \textbf{0.489} & \textbf{0.475} & \textbf{0.519} & \textbf{0.497} & 0.436 & \textbf{0.648} & \textbf{0.435} & 0.270 & \multirow{2}{*}{+15.1\%}\\
\quad \textit{w/o MAS} & 0.425 & 0.373 & 0.458 & 0.392 & \textbf{0.492} & 0.530 & 0.375 & \textbf{0.353} \\

\bottomrule
\end{tabular}
}
\vspace{-0.2cm}
\caption{\small Overall performance on \benchmark. Results are reported for each vanilla LLM and its \framework-enhanced counterparts. The best overall result is shown in red, the best standalone LLM result is \underline{underlined}. For each model pair, \textbf{bold} indicates the better result between the vanilla LLM and its \framework-enhanced counterpart. ``Improve'' denotes the relative final score gain of \framework\ over the corresponding vanilla LLM.
}

\label{tab:main_results}
\end{table*}

\subsection{Overall Results}
Table~\ref{tab:main_results} reports the experimental results. All \framework-enhanced LLMs consistently outperform corresponding vanilla LLMs in terms of final score, and excel corresponding LLMs on most sub-metrics (75 out of 84). \framework~(GPT-5.1) has the best performance on final score and most sub-metrics (4 out of 7). \framework-enhanced Claude family has the best model type selection and group definition score, while \framework~(Gemini-2.5-Pro) has the best core independent score. These results confirm the effectiveness of the \framework. 

We observe that the explanation and control metrics do not consistently improve with MAS. A likely contributing factor is simulation-induced selection bias, as the requirement for executable code encourages MAS to favor minimal, testable specifications. This tendency stabilizes the structural components of the design but may limit covariate coverage and reduce explanatory coherence.

For direct inference (vanilla LLMs \textit{w/o MAS}), GPT-5.1 achieves the highest score, best overall performance (6 out of 7), but relatively low score (0.493) on model type selection. In contrast, Claude and Gemini families perform better in model type selection and group definition. No single model dominates across all indicators, suggesting that end-to-end social-science causal study design remains challenging for current LLMs.

Even the strongest vanilla LLMs still exhibit limited performance in model type selection, group definition, reasoning and explanation (lower than 0.6). This result confirms LLMs function primarily as sophisticated semantic extractors rather than causal study designers.
They successfully retrieve relevant policy entities but frequently fail to synthesize these into a coherent identification strategy that is theoretically compatible with the data structure.

Beyond the main results, we conduct a series of supplementary analyses to validate the robustness and generalizability of our findings. We evaluate test-time scaling via best-of-$k$ sampling (Appendix~\ref{app_sec:best_of_k}), verify rubric sensitivity through re-weighting analysis (Appendix~\ref{app_sec:reweighting}), and confirm statistical significance across five independent runs (Appendix~\ref{app_sec:significance}). We also compare SFT-based approaches using both \benchmark\ and alternative training datasets, finding that SFT yields limited gains relative to MAS (Appendix~\ref{app_sec:sft_analysis}). Detailed per-method and per-domain error analyses are provided in Appendix~\ref{app_sec:error_analysis}, and full results for all 12 models under extended inference settings are reported in Appendix~\ref{app_sec:stronger_baselines}. These analyses collectively demonstrate that the performance gains of \framework\ are robust to evaluation design choices and are not attributable to naive test-time compute scaling.

\subsection{Cross-Model Extraction Consistency}
\label{app_sec:extraction_consistency}

To provide an objective, quantitative measure of annotation reliability, we conduct a cross-model extraction consistency check. Two state-of-the-art LLMs (GPT-5.1 and Gemini-3-Pro) independently extract the same fields directly from the source PDFs, and we compute field-wise agreement and Cohen's Kappa. As shown in Table~\ref{tab:kappa}, all key fields exhibit high agreement (83.1\%--96.4\%) with substantial to near-perfect Kappa (0.663--0.928). Structured fields such as Model Type , Control Variables, and Core Independent Variable achieve especially strong consistency.

\begin{table}[H]
\centering
\small
\setlength{\tabcolsep}{8pt}
\renewcommand{\arraystretch}{1.08}
\begin{tabular}{lcc}
\toprule
\textbf{Field} & \textbf{Agreement} & \textbf{Cohen's $\kappa$} \\
\midrule
Reasoning & 85.50\% & 0.790 \\
Core IV       & 91.60\% & 0.831 \\
Group Def                           & 89.20\% & 0.783 \\
Controls                & 94.00\% & 0.880 \\
Dep Var               & 83.10\% & 0.663 \\
Explanation                      & 84.30\% & 0.687 \\
Model type                       & 96.40\% & 0.928 \\
\bottomrule
\end{tabular}
\caption{Cross-model extraction consistency between GPT-5.1 and Gemini-3-Pro.}
\label{tab:kappa}
\end{table}

\subsection{Case Study: Benchmark Construction}
\label{app_sec:case_study}

This case study illustrates why fully automated multi-agent extraction is insufficient for constructing a research-grade causal benchmark and motivates the inclusion of a HITL adjudication stage.

\begin{figure}[H]
    \centering
    \includegraphics[width=0.9\columnwidth]{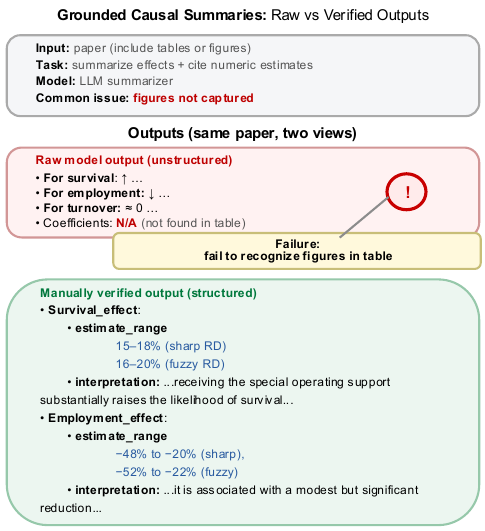}
    \caption{Case Study of Generating \benchmark. It presents a comparison between the raw output derived from the paper \citep{hammarlund2025impact} generated by our original LLM-based multi-agent system and the final output refined through human expert validation.}
    \label{fig:case study}
\end{figure}

As shown in Figure~\ref{fig:case study}, the initial outputs produced by our LLM-based multi-agent system fail to recover complete experimental results from long-context academic papers, particularly when key statistical evidence is distributed across multi-page tables or embedded in dense numerical appendices.

To address this limitation, we incorporate domain experts to validate and refine the extracted content before final inclusion in \benchmark. The HITL process focuses on verifying causal design completeness and resolving ambiguities that cannot be reliably adjudicated through automated reasoning alone. As illustrated in Figure~\ref{fig:case study}, expert adjudication yields a coherent and structured representation of the full experimental evidence, substantially improving both the completeness and reliability of the benchmark. 

\subsection{Ablation Study}
We implement three variants: \textbf{\textit{w/o MAS}} directly infers without MAS \framework, \textbf{Decomposition} only enables the theory architect agent and methodology agent, and \textbf{w/o Critic} disables critic agents.
We select GPT-5.1, Claude-Sonnet-4, and Gemini-2.5-Pro as representative case studies due to their superior overall performance. 

As Table~\ref{tab:ablation_comparison} shows, all modules enhance model performance. The theory architect agent and methodology agent contribute more to GPT-5.1 than the other two LLMs, suggesting that GPT-5.1 benefits more from explicit noise filtering and salient information extraction.
By contrast, the critic agent has a larger impact on GPT-5.1 and Claude-Sonnet-4, but a smaller effect on Gemini-2.5-Pro, indicating that Gemini-2.5-Pro exhibits stronger intrinsic self-correction based on code execution feedback. Full experimental results and analysis are detailed in Appendix~\ref{app_sec:ablation_results}.

\begin{table}[!t]
\centering
\small
\setlength{\tabcolsep}{1.5pt} 
\resizebox{\linewidth}{!}{
    \begin{tabular}{l|c|c|c}
    \toprule
    \textbf{Config} & \textbf{GPT-5.1} & \textbf{Claude-Sonnet-4} & \textbf{Gemini-2.5-Pro} \\
    \midrule
    \textit{w/o MAS} & 0.578 & 0.539 & 0.497 \\
    \midrule
    \multirow{2}{*}{\textit{Decomposition}} & 0.617 & 0.544 & 0.517 \\
     & \scriptsize{(\textcolor{teal}{+0.039})} & \scriptsize{(\textcolor{teal}{+0.005})} & \scriptsize{(\textcolor{teal}{+0.020})} \\
    \midrule
    \multirow{2}{*}{\textit{w/o Critic}} & 0.623 & 0.550 & 0.590 \\
     & \scriptsize{(\textcolor{teal}{+0.006})} & \scriptsize{(\textcolor{teal}{+0.006})} & \scriptsize{(\textbf{\textcolor{teal}{+0.073}})} \\
    \midrule
    \multirow{2}{*}{\textit{\framework}} & \textbf{0.665} & \textbf{0.652} & \textbf{0.621} \\
     & \scriptsize{(\textbf{\textcolor{teal}{+0.043}})} & \scriptsize{(\textbf{\textcolor{teal}{+0.102}})} & \scriptsize{(\textcolor{teal}{+0.031})} \\
    \bottomrule
    \end{tabular}
}
\vspace{-0.2cm}
\caption{\small Comparative Ablation Study. We report the overall score and the marginal gain (in parentheses).}
\label{tab:ablation_comparison}
\end{table}



\section{Conclusion}
This paper introduces \benchmark, a benchmark designed to evaluate causal reasoning as an end-to-end, intervention-centered and structure-agnostic research process in real-world social science settings. By grounding evaluation in empirical policy interventions and requiring models to reason under implicit assumptions and open-ended contexts, \benchmark\ reveals substantial gaps between the capabilities of current large language models and the demands of research-grade causal inquiry.
Our empirical results show that even SOTA LLMs struggle to construct coherent causal study designs when faced with realistic interventions, despite strong performance on existing reasoning benchmarks. To illustrate a potential path forward, we present \framework, a multi-agent framework that partially mitigates these limitations by structuring causal reasoning around expert-inspired roles.
We hope \benchmark\ will serve as a foundation for more faithful evaluation of causal reasoning in natural language models and encourage future work on developing LLM systems capable of supporting rigorous, real-world social science research.

\section*{Limitations}
This study has some limitations. \benchmark\ is constructed from peer-reviewed empirical social science studies and therefore reflects the scope and availability of existing policy-oriented research, potentially underrepresenting social studies with limited empirical coverage. The benchmark focuses on evaluating research-design reasoning rather than the numerical accuracy of causal effect estimation, and thus does not capture errors arising from downstream statistical implementation. In addition, the evaluation abstracts complex causal study designs into a set of discrete criteria, which may rely on simulation-based verification and external tools that cannot fully capture real-world data complexities.

\section*{Ethical considerations}
Our dataset construction includes validation by four domain experts to ensure the correctness of policy descriptions and causal study specifications. All expert work was conducted in compliance with applicable local laws and regulations, and experts were compensated for their labor. Each sample was cross-checked by at least two experts. The validation tasks did not require experts to provide sensitive personal information, and the dataset is derived from publicly available academic publications. We did not recruit human subjects for behavioral studies, and we did not collect or release any personally identifiable information.



\bibliography{custom}

\newpage
\appendix

\section*{Appendix}


\section{Additional Experimental Details}
\subsection{Detailed Benchmark}
\label{app_sec:benchmark}
\benchmark\ consists of 744 empirical social science studies, each corresponding to a policy intervention and its associated causal study design. During instance generating processing,  36 out of 744 initial outputs from the MAS fails to capture the complete statistical results embedded, and refined through rigorous HITL adjudication.

\subsection{Detailed Experimental Setup}
\label{app_sec:metric}
\paragraph{Implementation.} 
We adopt a hierarchical rubric spanning three dimensions: \emph{Core Causal Design}, \emph{Model Specification}, and \emph{Logic \& Interpretation}. The total raw score is 45 points, which is normalized to a $[0, 1]$ scale for the final report. InterveneBench is scored with a component-wise, graded rubric that measures how closely a predicted study design matches the expert-verified ground truth (allowing partial credit for partially aligned components), rather than treating any single mistake in the final outcome as an all-or-nothing failure. Standard details are as follows:

\emph{Core Causal Design (30 pts).}
This dimension assesses the structural validity of the proposed identification framework.

-- \textit{Model Type (10 pts):} The most critical metric. It evaluates whether the model selects the correct causal inference method (e.g., DiD, IV, RDD) given the policy context. Scoring is binary: full marks are awarded only for a strict semantic match with the expert-annotated ground truth (e.g., "Difference-in-Differences" matches "DiD"); otherwise, 0 points.

-- \textit{Core Independent Variable (10 pts):} Evaluates the operationalization of the policy intervention. 10 points for precise alignment with the ground truth; 5 points if the primary features are captured but vaguely described; 0 points for incorrect identification.

-- \textit{Group Definition (10 pts):} Assesses the delineation of the \textit{Treatment} and \textit{Control} groups. 10 points for a clear and accurate definition of the affected population versus the counterfactual sample; 5 points for ambiguous definitions.

\emph{Model Auxiliary Details (10 pts).}
This dimension measures the completeness of the statistical specification.

--\textit{Control Variables (5 pts):} Evaluates the model's ability to account for confounders (e.g., economic indicators, demographics). 5 points if the coverage of ground truth variables exceeds 50\%; 2 points for partial coverage.

--\textit{Dependent Variable (5 pts):} Assesses the identification of the target outcome variable intended to be influenced by the policy. 5 points for semantic consistency with the ground truth.

\emph{Logic \& Interpretation (5 pts).}
This dimension evaluates the depth and consistency of the scientific reasoning.

--\textit{Reasoning (2 pts):} Examines the logical self-consistency of the methodological choice. 

--\textit{Explanation (3 pts):} Evaluates whether the predicted direction of the policy effect and the proposed mechanism align with  social science academic consensus.

\begin{table*}[!t]
\centering
\small
\setlength{\tabcolsep}{5pt}
\renewcommand{\arraystretch}{1.08}
\resizebox{\textwidth}{!}{
\begin{tabular}{lcccccccc|c}
\toprule
Model & Final Score & Model Type & Core IV & Group Def & Controls & Dep Var & Reasoning & Explanation & Improve \\
\midrule

STRIDES(gpt-5.1) & 0.665 & 0.515 & 0.691 & 0.544 & 0.842 & 0.974 & 0.695 & 0.650 & {+15.1\%} \\
\quad \textit{Decomposition} & 0.617 & 0.540 & 0.631 & 0.619 & 0.638 & 0.820 & 0.540 & 0.497 & {+6.8\%} \\
\quad \textit{w/o critic} & 0.623 & 0.527 & 0.654 & 0.594 & 0.634 & 0.870 & 0.525 & 0.570 & {+7.8\%} \\
\quad \textit{w/o MAS} & 0.578 & 0.493 & 0.656 & 0.535 & 0.652 & 0.658 & 0.530 & 0.517 & {-} \\
\midrule

STRIDES(gpt-4.1) & 0.570 & 0.541 & 0.646 & 0.580 & 0.430 & 0.756 & 0.540 & 0.320 & {+25.0\%} \\
\quad \textit{Decomposition} & 0.501 & 0.455 & 0.568 & 0.518 & 0.346 & 0.794 & 0.455 & 0.183 & {+10.0\%} \\
\quad \textit{w/o critic} & 0.542 & 0.524 & 0.614 & 0.531 & 0.400 & 0.800 & 0.525 & 0.220 & {+19.0\%} \\
\quad \textit{w/o MAS} & 0.456 & 0.445 & 0.498 & 0.430 & 0.458 & 0.572 & 0.445 & 0.243 & {-} \\
\midrule

STRIDES(gemini-2.5-pro) & 0.621 & 0.646 & 0.762 & 0.578 & 0.500 & 0.740 & 0.600 & 0.233 & {+25.0\%} \\
\quad \textit{Decomposition} & 0.517 & 0.473 & 0.573 & 0.519 & 0.368 & 0.748 & 0.455 & 0.380 & {+4.1\%} \\
\quad \textit{w/o critic} & 0.590 & 0.526 & 0.631 & 0.582 & 0.474 & 0.884 & 0.525 & 0.433 & {+18.6\%} \\
\quad \textit{w/o MAS} & 0.497 & 0.477 & 0.509 & 0.477 & 0.508 & 0.594 & 0.475 & 0.427 & {-} \\
\midrule

STRIDES(grok-4) & 0.580 & 0.550 & 0.657 & 0.591 & 0.436 & 0.770 & 0.550 & 0.327 & {+25.0\%} \\
\quad \textit{Decomposition} & 0.554 & 0.544 & 0.569 & 0.544 & 0.392 & 0.742 & 0.540 & 0.530 & {+19.3\%} \\
\quad \textit{w/o critic} & 0.573 & 0.495 & 0.665 & 0.537 & 0.366 & 0.818 & 0.475 & 0.643 & {+23.4\%} \\
\quad \textit{w/o MAS} & 0.464 & 0.424 & 0.514 & 0.445 & 0.462 & 0.572 & 0.420 & 0.340 & {-} \\
\midrule

STRIDES(gemini-3-flash-preview) & 0.583 & 0.525 & 0.638 & 0.586 & 0.460 & 0.750 & 0.530 & 0.547 & {+25.0\%} \\
\quad \textit{Decomposition} & 0.512 & 0.492 & 0.552 & 0.510 & 0.354 & 0.648 & 0.495 & 0.503 & {+9.9\%} \\
\quad \textit{w/o critic} & 0.544 & 0.513 & 0.590 & 0.531 & 0.400 & 0.692 & 0.515 & 0.547 & {+16.6\%} \\
\quad \textit{w/o MAS} & 0.466 & 0.445 & 0.477 & 0.477 & 0.434 & 0.552 & 0.445 & 0.393 & {-} \\
\midrule

STRIDES(claude-sonnet-4) & 0.652 & 0.634 & 0.734 & 0.652 & 0.550 & 0.816 & 0.615 & 0.367 & {+21.1\%} \\
\quad \textit{Decomposition} & 0.544 & 0.454 & 0.564 & 0.542 & 0.560 & 0.776 & 0.425 & 0.453 & {+1.0\%} \\
\quad \textit{w/o critic} & 0.550 & 0.467 & 0.573 & 0.551 & 0.540 & 0.738 & 0.450 & 0.513 & {+2.0\%} \\
\quad \textit{w/o MAS} & 0.539 & 0.541 & 0.573 & 0.541 & 0.488 & 0.656 & 0.530 & 0.307 & {-} \\
\midrule

STRIDES(claude-3-7-sonnet) & 0.653 & 0.650 & 0.728 & 0.601 & 0.570 & 0.892 & 0.620 & 0.347 & {+19.9\%} \\
\quad \textit{Decomposition} & 0.546 & 0.509 & 0.606 & 0.502 & 0.454 & 0.718 & 0.510 & 0.510 & {+0.3\%} \\
\quad \textit{w/o critic} & 0.554 & 0.505 & 0.600 & 0.513 & 0.464 & 0.726 & 0.470 & 0.617 & {+1.7\%} \\
\quad \textit{w/o MAS} & 0.544 & 0.519 & 0.578 & 0.535 & 0.570 & 0.656 & 0.510 & 0.340 & {-} \\
\midrule

STRIDES(deepseek-v3.2) & 0.489 & 0.475 & 0.519 & 0.497 & 0.436 & 0.648 & 0.435 & 0.270 & {+15.0\%} \\
\quad \textit{Decomposition} & 0.436 & 0.417 & 0.448 & 0.423 & 0.390 & 0.614 & 0.390 & 0.307 & {+2.4\%} \\
\quad \textit{w/o critic} & 0.444 & 0.413 & 0.454 & 0.424 & 0.366 & 0.656 & 0.390 & 0.400 & {+4.5\%} \\
\quad \textit{w/o MAS} & 0.425 & 0.373 & 0.458 & 0.392 & 0.492 & 0.530 & 0.375 & 0.353 & {-} \\
\midrule

STRIDES(Kimi-K2-0905) & 0.519 & 0.523 & 0.555 & 0.484 & 0.424 & 0.674 & 0.480 & 0.420 & {+25.0\%} \\
\quad \textit{Decomposition} & 0.428 & 0.424 & 0.450 & 0.370 & 0.360 & 0.582 & 0.395 & 0.437 & {+3.2\%} \\
\quad \textit{w/o critic} & 0.468 & 0.417 & 0.500 & 0.462 & 0.376 & 0.702 & 0.415 & 0.343 & {+12.7\%} \\
\quad \textit{w/o MAS} & 0.415 & 0.361 & 0.445 & 0.414 & 0.438 & 0.508 & 0.360 & 0.340 & {-} \\
\midrule

STRIDES(glm-4.6) & 0.652 & 0.613 & 0.722 & 0.657 & 0.470 & 0.898 & 0.610 & 0.460 & {+21.1\%} \\
\quad \textit{Decomposition} & 0.544 & 0.585 & 0.533 & 0.523 & 0.482 & 0.608 & 0.565 & 0.503 & {+1.0\%} \\
\quad \textit{w/o critic} & 0.605 & 0.562 & 0.732 & 0.550 & 0.398 & 0.782 & 0.560 & 0.587 & {+12.2\%} \\
\quad \textit{w/o MAS} & 0.539 & 0.535 & 0.566 & 0.533 & 0.480 & 0.644 & 0.520 & 0.303 & {-} \\
\midrule

STRIDES(gpt-oss-120b) & 0.553 & 0.553 & 0.611 & 0.512 & 0.436 & 0.726 & 0.550 & 0.413 & {+25.0\%} \\
\quad \textit{Decomposition} & 0.445 & 0.438 & 0.484 & 0.421 & 0.320 & 0.586 & 0.415 & 0.413 & {+0.5\%} \\
\quad \textit{w/o critic} & 0.502 & 0.530 & 0.530 & 0.443 & 0.312 & 0.716 & 0.500 & 0.470 & {+13.4\%} \\
\quad \textit{w/o MAS} & 0.443 & 0.435 & 0.440 & 0.403 & 0.466 & 0.572 & 0.415 & 0.377 & {-} \\
\midrule

STRIDES(qwen-235b-a22b) & 0.569 & 0.538 & 0.638 & 0.544 & 0.534 & 0.712 & 0.525 & 0.373 & {+18.0\%} \\
\quad \textit{Decomposition} & 0.506 & 0.479 & 0.568 & 0.484 & 0.476 & 0.634 & 0.465 & 0.333 & {+5.0\%} \\
\quad \textit{w/o critic} & 0.531 & 0.502 & 0.595 & 0.508 & 0.498 & 0.664 & 0.490 & 0.350 & {+10.0\%} \\
\quad \textit{w/o MAS} & 0.482 & 0.456 & 0.541 & 0.461 & 0.452 & 0.604 & 0.445 & 0.317 & {-} \\

\bottomrule
\end{tabular}
}
\caption{\small Ablation results on \benchmark. ``Improve'' denotes the relative final score gain of variants of \framework\ over the corresponding vanilla LLM (w/o MAS).}
\label{tab:ablation_big}
\end{table*}

\subsection{Ablation Study}
\label{app_sec:ablation_results}
We implement three variants: \textbf{\textit{w/o MAS}} directly infers without MAS \framework, \textbf{Decomposition} only enables the theory architect agent and methodology agent, and \textbf{w/o Critic} disables critic agents.Table~\ref{tab:ablation_big} presents the complete ablation experiment results. As results show, all modules enhance model performance, confirming the effectiveness of these modules.

The \textit{Decomposition} setting consistently improves over \textit{w/o MAS}. Within the GPT family, GPT-5.1 and GPT-4.1 benefit more from the Decomposition setting than GPT-OSS-120B, suggesting that stronger models more effectively leverage decomposed intermediate representations. By comparison, Claude models show consistently smaller gains, indicating that explicit decomposition provides more limited additional benefit for this model family.

By contrast, the critic agent has a larger impact on the Claude-Sonnet family and DeepSeek-V3.2, suggesting more limited intrinsic self-correction from code execution feedback for these models.

\section{Case Study: LLM-Judge Consistency}
\label{app_case_con}
We conduct an LLM-judge robustness case study on a 5\% random subset of the test set. Domain experts rewrite each sampled instance to alter the surface form while preserving its core semantics. We then re-score the rewritten instances using the same Grader (Gemini-3-Pro)\cite{googledeepmind2025gemini3pro_modelcard} and ask the experts to verify the outcomes. Results indicate that the scores are stable under paraphrasing, and the core metrics remain consistent with expert judgments. Table~\ref{tab:rewriting_comparison} shows a representative example.
\begin{table*}[!t]
\centering
\small
\begin{tabularx}{\textwidth}{>{\raggedright\arraybackslash}X >{\raggedright\arraybackslash}X >{\raggedright\arraybackslash}X >{\raggedright\arraybackslash}X}
\toprule
\textbf{} & \textbf{Original Ground-Truth} & \textbf{Rewritten Ground-Truth} & \textbf{Output} \\
\midrule
\textbf{Model Type} & 
{\footnotesize Regression Discontinuity (RD)} & 
{\footnotesize Regression Discontinuity (RD)} & 
{\footnotesize Sharp RD on clinical BP thresholds in panel setting}\\
\addlinespace[2ex]

\textbf{Core Independent Variable} & 
{\footnotesize First-ever hypertension diagnosis at wave t triggered by measured SBP $\geq$ 140 mmHg or DBP $\geq$ 90 mmHg at t-1.} & 
{\footnotesize First-ever hypertension diagnosis at wave t triggered by measured SBP $\geq$ 140 mmHg or DBP $\geq$ 90 mmHg at previous wave (t-1).} & 
{\footnotesize First-ever hypertension diagnosis at wave t based on: (i) measured SBP $\geq$ 140 mmHg or DBP $\geq$ 90 mmHg; (ii) clinician diagnosis by wave t; (iii) no prior diagnosis. RD uses distance from thresholds as running variable.} \\
\addlinespace[2ex]

\textbf{Group} & 
{\footnotesize 
\textbf{Definition:} First-ever
diagnosis at t triggered by BP at t-1 crossing thresholds.
\textbf{Key RD sets (SBP-based):}
- \textit{Treatment:} DBP $<$ 90, SBP $\geq$ 140 (local), newly diagnosed.
- \textit{Control:} DBP $<$ 90, SBP $<$ 140 (local), undiagnosed.
- \textit{Both:} No prior diagnosis; outcomes at t ($\sim$3 years later).
\textbf{DBP-based RD:}
- \textit{Treatment:} SBP $<$ 140, DBP $\geq$ 90, newly diagnosed.
- \textit{Control:} SBP $<$ 140, DBP $<$ 90, undiagnosed.
\textbf{Distance-to-boundary RD:}
- \textit{Treatment:} Just on diagnosed side of (SBP, DBP) boundary.
- \textit{Control:} Just on non-diagnosed side.
} & 
{\footnotesize 
\textbf{Definition:} Initial diagnosis at t triggered by BP at t-1 crossing thresholds.
\textbf{Primary RD (SBP-based):}
- \textit{Treatment:} DBP $<$ 90, SBP $\geq$ 140 (local), newly diagnosed.
- \textit{Control:} DBP $<$ 90, SBP $<$ 140 (local), undiagnosed.
- \textit{Restrictions:} No prior diagnosis; outcomes at t ($\sim$3 years later).
\textbf{DBP-based RD:}
- \textit{Treatment:} SBP $<$ 140, DBP $\geq$ 90, newly diagnosed.
- \textit{Control:} SBP $<$ 140, DBP $<$ 90, undiagnosed.
\textbf{Distance-to-boundary RD:}
- \textit{Treatment:} Just on diagnosed side of boundary.
- \textit{Control:} Just on non-diagnosed side.
} & 
{\footnotesize 
\textbf{Treatment:} CHNS adults (age $\geq$ 18) with first-time measured SBP $\geq$ 140 or DBP $\geq$ 90 at exam wave, receiving first-ever diagnosis at that 'teachable moment'. RD sample restricted to bandwidth around cutoff(s).
\textbf{Control:} Comparable adults with BP just below both thresholds (e.g., SBP in [140 $-$ h, 140) and DBP in [90 $-$ h, 90)), never previously diagnosed. They serve as local counterfactual for diet evolution absent informational shock.
} \\
\bottomrule
\end{tabularx}
\caption{Comparison among original benchmark, rewritten benchmark and vanilla LLM's output}
\label{tab:rewriting_comparison}
\end{table*}


\section{SFT and Legacy Set Analysis}
\label{app_sec:sft_analysis}

\subsection{Results on Legacy Set}
\label{app:result_lag}
Table~\ref{app:legacy_gpt51} presents the results on the Legacy Set. GPT-5.1's performance on the Legacy Set is significantly higher than those on the Test Set; while the additional performance gain (+5.83\%) from the \framework\ framework becomes smaller. This pattern likely reflects the presence of these documents in the pre-training corpus, indicating that the models are retrieving memorized ground truths rather than generalizing the underlying logic of causal inference.

\begin{table}[!t]
\centering
\small
\setlength{\tabcolsep}{6pt}
\renewcommand{\arraystretch}{1.15}
\begin{tabular}{lcc}
\toprule
Metric & \textit{w/o MAS} & \framework \\
\midrule
Final Score  & 0.7456 & 0.7891 \\
Model Type   & 0.5290 & 0.5600 \\
Core IV      & 0.8280 & 0.8770 \\
Group Def    & 0.7260 & 0.7680 \\
Controls     & 0.9420 & 0.9960 \\
Dep Var      & 0.9240 & 0.9780 \\
Reasoning    & 0.5250 & 0.5550 \\
Explanation  & 0.7800 & 0.8267 \\
\bottomrule
\end{tabular}
\caption{\small GPT-5.1 results on the Legacy Set.}
\label{app:legacy_gpt51}
\end{table}

\subsection{Results after Legacy-Set SFT}
We fine-tune the open-source model Qwen3-32B \cite{yang2025qwen3} using the LLaMA-Factory \citep{zheng2024llamafactory} framework on the Legacy-Set dataset. The results are shown in Table~\ref{tab:sft_comparison}. After SFT, direct inference performance improves by 7.6\%. However, under our framework setting, the SFT model performs almost the same as the base model. We attribute this diminishing return to the diverse functional requirements of agents within our MAS. SFT appears to induce catastrophic forgetting and reduce the model's generalizability \citep{li-etal-2024-revisiting,wang-etal-2024-inscl,gupta-etal-2025-selective}. Consequently, the model loses the flexibility required for specialized agent roles (e.g., critique or simulation), substantially dampening the performance benefits typically provided by the MAS.

\subsection{SFT with Alternative Training Datasets}
\label{app_sec:sft_baseline}

To investigate whether performance gains stem from the training data itself, we fine-tune Qwen3-32B using CauSciBench~\citep{acharya2025causcibench} as an alternative training set. As shown in Table~\ref{tab:sft_comparison}, results are essentially unchanged across SFT sources (e.g., STRIDES: 0.5493 vs.\ 0.5496), while the dominant gap remains the presence versus absence of MAS. This suggests that SFT provides only limited incremental gains; the main bottleneck is inducing robust reasoning behaviors~\citep{guo2025deepseek}.

\begin{table*}[t]
\centering
\small
\setlength{\tabcolsep}{5pt}
\renewcommand{\arraystretch}{1.0}
\resizebox{1.0\textwidth}{!}{%
\begin{tabular}{lcccccccc}
\toprule
\textbf{Model} & \textbf{Final Score} & \textbf{Model Type} & \textbf{Core IV} & \textbf{Group Def} & \textbf{Controls} & \textbf{Dep Var} & \textbf{Reasoning} & \textbf{Explanation} \\
\midrule
STRIDES (base\_model)              & 0.5493 & 0.5010 & 0.5750 & 0.5260 & 0.4900 & 0.7380 & 0.4900 & 0.5267 \\
\textit{w/o MAS} (base\_model)     & 0.4653 & 0.4240 & 0.4880 & 0.4450 & 0.4160 & 0.6240 & 0.4150 & 0.4467 \\
\midrule
STRIDES (CauSciBench)              & 0.5496 & 0.5029 & 0.5753 & 0.5260 & 0.4941 & 0.7334 & 0.4947 & 0.5214 \\
\textit{w/o MAS} (CauSciBench)     & 0.4660 & 0.4294 & 0.4947 & 0.4505 & 0.4222 & 0.6303 & 0.4197 & 0.3766 \\
\midrule
STRIDES (InterveneBench)           & 0.5548 & 0.5040 & 0.5771 & 0.5283 & 0.4953 & 0.7424 & 0.5208 & 0.5479 \\
\textit{w/o MAS} (InterveneBench)  & 0.5007 & 0.4820 & 0.5720 & 0.4580 & 0.4380 & 0.6260 & 0.4800 & 0.3730 \\
\bottomrule
\end{tabular}%
}
\caption{Comparison of SFT with different training sources (Qwen3-32B). The dominant gap remains between STRIDES and \textit{w/o MAS}, not between SFT sources.}
\label{tab:sft_comparison}
\end{table*}

\section{Best-of-\textit{k} Test-Time Scaling Analysis}
\label{app_sec:best_of_k}

We evaluate a common test-time scaling heuristic: running $k$ independent generations with Kimi-K2 and selecting the best output. As shown in Table~\ref{tab:best_of_k}, gains saturate quickly. Under a token budget comparable to MAS, best-of-$k$ yields only 7.54\% improvement, whereas MAS achieves $\sim$25\%.

\begin{table}[H]
\centering
\small
\setlength{\tabcolsep}{10pt}
\renewcommand{\arraystretch}{1.08}
\begin{tabular}{ccc}
\toprule
\textbf{$k$} & \textbf{Avg.\ Final Score} & \textbf{$\Delta$ vs.\ $k{-}1$} \\
\midrule
1 & 0.4150 & ---   \\
2 & 0.4249 & +2.38\% \\
3 & 0.4397 & +3.49\% \\
4 & 0.4419 & +0.50\% \\
5 & 0.4419 & 0\%    \\
6 & 0.4463 & +1.00\% \\
\bottomrule
\end{tabular}
\caption{Best-of-$k$ test-time scaling with Kimi-K2.}
\label{tab:best_of_k}
\end{table}

\section{Robustness and Statistical Significance}
\label{app_sec:reweighting}

\subsection{Re-Weighting Sensitivity}

We increase the weight of Logic \& Interpretation from 5 to 15 points and Model Specification from 10 to 15 points (new total: 60 points, re-normalized to $[0,1]$). As shown in Table~\ref{tab:reweight}, overall rankings remain largely stable, confirming that findings reflect consistent differences in causal design competence.

\begin{table}[H]
\centering
\small
\setlength{\tabcolsep}{8pt}
\renewcommand{\arraystretch}{1}
\resizebox{0.5\textwidth}{!}{
\begin{tabular}{lcc}
\toprule
\textbf{Model} & \textbf{Rank $\Delta$} & \textbf{Final Score} \\
\midrule
STRIDES (GPT-5.1)            & --    & 0.828 \\
STRIDES (Claude-3.7-Sonnet)  & --    & 0.639 \\
STRIDES (GLM-4.6)            & $\uparrow$1  & 0.619 \\
STRIDES (Claude-Sonnet-4)    & $\downarrow$1  & 0.611 \\
STRIDES (Gemini-3-Flash)     & $\uparrow$1  & 0.583 \\
STRIDES (Qwen3-235B-A22B)    & $\uparrow$3  & 0.560 \\
STRIDES (GPT-OSS-120B)       & $\uparrow$3  & 0.543 \\
STRIDES (Grok-4)             & $\downarrow$1  & 0.541 \\
STRIDES (Gemini-2.5-Pro)     & $\downarrow$4  & 0.540 \\
STRIDES (GPT-4.1)            & $\downarrow$2  & 0.531 \\
STRIDES (Kimi-K2)            & --    & 0.514 \\
STRIDES (DeepSeek-v3.2)      & --    & 0.473 \\
\bottomrule
\end{tabular}
}
\caption{Re-weighted scoring analysis.}
\label{tab:reweight}
\end{table}

\subsection{Statistical Significance and Stability}
\label{app_sec:significance}

\paragraph{Run-to-run stability.}
We evaluate the full test set five independent times using GPT-5.1 + STRIDES. The standard deviation across runs is 0.0126, with a 95\% confidence interval of [0.6498, 0.6762]. A Friedman test yields $p = 0.77$ ($> 0.05$), confirming no significant run-to-run variation.

\begin{table}[H]
\centering
\small
\setlength{\tabcolsep}{8pt}
\renewcommand{\arraystretch}{1.08}
\begin{tabular}{lc}
\toprule
\textbf{Run} & \textbf{Final Score} \\
\midrule
Run 1 & 0.6650 \\
Run 2 & 0.6580 \\
Run 3 & 0.6720 \\
Run 4 & 0.6510 \\
Run 5 & 0.6690 \\
\midrule
Mean $\pm$ Std & 0.6630 $\pm$ 0.0126 \\
95\% CI & [0.6498, 0.6762] \\
\bottomrule
\end{tabular}
\caption{Run-to-run stability of GPT-5.1 + STRIDES.}
\label{tab:stability}
\end{table}

\paragraph{Framework significance.}
The Vanilla mean is 0.5686 (95\% CI: [0.5539, 0.5833]), while the STRIDES mean is 0.6630 (95\% CI: [0.6498, 0.6762]). A paired $t$-test yields $p < 10^{-6}$, confirming statistical significance with non-overlapping confidence intervals.

\paragraph{Per-metric significance.}
Using a Wilcoxon signed-rank test across five paired runs, all core metrics show significant improvements ($p < 0.05$). The Control Variables metric does not reach significance ($p = 0.12$), consistent with our observation that MAS does not consistently improve covariate coverage.

\section{Full Results for Stronger Baselines}
\label{app_sec:stronger_baselines}

\paragraph{Baseline Descriptions.}
In addition to \textit{Vanilla} (direct zero-shot inference) and \textit{STRIDES} (our multi-agent framework), we evaluate three standard inference-time enhancement baselines:

\begin{itemize}
    \item \textbf{Few-shot (3-shot).} We prepend three expert-annotated exemplars (one DiD, one IV, one RD) to the prompt. Each exemplar contains the policy context and the corresponding ground-truth causal study design, providing in-context demonstrations of the expected output format and reasoning pattern.
    \item \textbf{CoT + Self-Refine.} The model first generates a chain-of-thought response, then explicitly critiques its own output and produces a refined answer in a second pass. This tests whether self-correction alone can approach the gains of multi-agent collaboration.
    \item \textbf{Best-of-6.} We generate six independent responses (temperature = 0.7) and select the one with the highest self-assessed confidence score. This serves as a test-time compute scaling baseline, consuming a token budget comparable to STRIDES ($\sim$20{,}160 tokens vs.\ $\sim$19{,}000 for STRIDES).
\end{itemize}

Table~\ref{tab:full_baselines} reports results for all 12 backbone LLMs under five inference settings. Across all backbones, the performance gradient is consistent: Few-shot yields 2--4\% gains, CoT + Self-Refine yields 5--7\%, Best-of-6 yields 6--9\%, while STRIDES achieves 15--25\%.

\begin{table*}[t]
\centering
\setlength{\tabcolsep}{4pt}
\adjustbox{max width=\textwidth, max totalheight=0.92\textheight}{
\begin{tabular}{ll|ccccccc|c}
\toprule
\textbf{Backbone} & \textbf{Setting} & \textbf{Final Score} & \textbf{Model Type} & \textbf{Core IV} & \textbf{Group Def} & \textbf{Controls} & \textbf{Dep Var} & \textbf{Reasoning} & \textbf{Explanation} \\
\midrule

\multirow{5}{*}{GPT-5.1}
& Vanilla      & 0.578 & 0.493 & 0.656 & 0.535 & 0.652 & 0.658 & 0.530 & 0.517 \\
& Few-shot     & 0.597 & 0.500 & 0.672 & 0.544 & 0.660 & 0.680 & 0.545 & 0.530 \\
& CoT+Refine   & 0.612 & 0.507 & 0.680 & 0.551 & 0.664 & 0.700 & 0.558 & 0.540 \\
& Best-of-6    & 0.620 & 0.510 & 0.685 & 0.548 & 0.668 & 0.708 & 0.555 & 0.545 \\
& \textbf{STRIDES} & \textbf{0.665} & \textbf{0.515} & \textbf{0.691} & \textbf{0.544} & \textbf{0.842} & \textbf{0.974} & \textbf{0.695} & \textbf{0.650} \\
\midrule

\multirow{5}{*}{Claude-3.7-Sonnet}
& Vanilla      & 0.544 & 0.519 & 0.578 & 0.535 & 0.570 & 0.656 & 0.510 & 0.340 \\
& Few-shot     & 0.561 & 0.528 & 0.596 & 0.546 & 0.577 & 0.674 & 0.524 & 0.352 \\
& CoT+Refine   & 0.577 & 0.537 & 0.610 & 0.554 & 0.582 & 0.690 & 0.535 & 0.360 \\
& Best-of-6    & 0.584 & 0.540 & 0.616 & 0.552 & 0.586 & 0.696 & 0.534 & 0.363 \\
& \textbf{STRIDES} & \textbf{0.653} & \textbf{0.650} & \textbf{0.728} & \textbf{0.601} & \textbf{0.570} & \textbf{0.892} & \textbf{0.620} & \textbf{0.347} \\
\midrule

\multirow{5}{*}{Claude-Sonnet-4}
& Vanilla      & 0.539 & 0.541 & 0.573 & 0.541 & 0.488 & 0.656 & 0.530 & 0.307 \\
& Few-shot     & 0.555 & 0.548 & 0.590 & 0.550 & 0.496 & 0.672 & 0.540 & 0.318 \\
& CoT+Refine   & 0.572 & 0.556 & 0.603 & 0.558 & 0.502 & 0.686 & 0.548 & 0.327 \\
& Best-of-6    & 0.580 & 0.562 & 0.608 & 0.555 & 0.506 & 0.690 & 0.545 & 0.330 \\
& \textbf{STRIDES} & \textbf{0.652} & \textbf{0.634} & \textbf{0.734} & \textbf{0.652} & \textbf{0.550} & \textbf{0.816} & \textbf{0.615} & \textbf{0.367} \\
\midrule

\multirow{5}{*}{GLM-4.6}
& Vanilla      & 0.531 & 0.535 & 0.566 & 0.533 & 0.480 & 0.644 & 0.520 & 0.303 \\
& Few-shot     & 0.548 & 0.543 & 0.582 & 0.544 & 0.488 & 0.662 & 0.533 & 0.315 \\
& CoT+Refine   & 0.563 & 0.552 & 0.596 & 0.553 & 0.492 & 0.678 & 0.544 & 0.323 \\
& Best-of-6    & 0.570 & 0.556 & 0.602 & 0.550 & 0.496 & 0.684 & 0.542 & 0.326 \\
& \textbf{STRIDES} & \textbf{0.642} & \textbf{0.606} & \textbf{0.709} & \textbf{0.646} & \textbf{0.464} & \textbf{0.882} & \textbf{0.600} & \textbf{0.453} \\
\midrule

\multirow{5}{*}{Gemini-2.5-Pro}
& Vanilla      & 0.497 & 0.477 & 0.509 & 0.477 & 0.508 & 0.594 & 0.475 & 0.427 \\
& Few-shot     & 0.513 & 0.486 & 0.525 & 0.488 & 0.514 & 0.612 & 0.488 & 0.437 \\
& CoT+Refine   & 0.530 & 0.495 & 0.540 & 0.498 & 0.518 & 0.628 & 0.498 & 0.440 \\
& Best-of-6    & 0.537 & 0.500 & 0.546 & 0.500 & \textbf{0.520} & 0.635 & 0.500 & 0.442 \\
& \textbf{STRIDES} & \textbf{0.621} & \textbf{0.646} & \textbf{0.762} & \textbf{0.578} & 0.500 & \textbf{0.740} & \textbf{0.600} & 0.233 \\
\midrule

\multirow{5}{*}{Gemini-3-Flash}
& Vanilla      & 0.466 & 0.445 & 0.477 & 0.477 & 0.434 & 0.552 & 0.445 & 0.393 \\
& Few-shot     & 0.482 & 0.455 & 0.492 & 0.488 & 0.441 & 0.570 & 0.458 & 0.403 \\
& CoT+Refine   & 0.496 & 0.464 & 0.505 & 0.497 & 0.446 & 0.586 & 0.469 & 0.410 \\
& Best-of-6    & 0.502 & 0.468 & 0.510 & 0.496 & 0.448 & 0.592 & 0.468 & 0.414 \\
& \textbf{STRIDES} & \textbf{0.583} & \textbf{0.525} & \textbf{0.638} & \textbf{0.586} & \textbf{0.460} & \textbf{0.750} & \textbf{0.530} & \textbf{0.547} \\
\midrule

\multirow{5}{*}{Grok-4}
& Vanilla      & 0.464 & 0.424 & 0.514 & 0.445 & 0.462 & 0.572 & 0.420 & 0.340 \\
& Few-shot     & 0.480 & 0.434 & 0.528 & 0.456 & 0.468 & 0.590 & 0.434 & 0.352 \\
& CoT+Refine   & 0.494 & 0.444 & 0.540 & 0.465 & 0.473 & 0.604 & 0.446 & 0.360 \\
& Best-of-6    & 0.500 & 0.448 & 0.546 & 0.463 & \textbf{0.476} & 0.610 & 0.444 & 0.364 \\
& \textbf{STRIDES} & \textbf{0.580} & \textbf{0.550} & \textbf{0.657} & \textbf{0.591} & 0.436 & \textbf{0.770} & \textbf{0.550} & 0.327 \\
\midrule

\multirow{5}{*}{GPT-4.1}
& Vanilla      & 0.456 & 0.445 & 0.498 & 0.430 & 0.458 & 0.572 & 0.445 & 0.243 \\
& Few-shot     & 0.470 & 0.454 & 0.513 & 0.441 & 0.464 & 0.588 & 0.456 & 0.254 \\
& CoT+Refine   & 0.485 & 0.462 & 0.526 & 0.451 & 0.469 & 0.603 & 0.466 & 0.262 \\
& Best-of-6    & 0.491 & 0.466 & 0.532 & 0.449 & \textbf{0.472} & 0.608 & 0.465 & 0.265 \\
& \textbf{STRIDES} & \textbf{0.570} & \textbf{0.541} & \textbf{0.646} & \textbf{0.580} & 0.430 & \textbf{0.756} & \textbf{0.540} & \textbf{0.320} \\
\midrule

\multirow{5}{*}{Qwen3-235B-A22B}
& Vanilla      & 0.482 & 0.456 & 0.541 & 0.461 & 0.452 & 0.604 & 0.445 & 0.317 \\
& Few-shot     & 0.497 & 0.465 & 0.556 & 0.472 & 0.458 & 0.620 & 0.457 & 0.328 \\
& CoT+Refine   & 0.511 & 0.474 & 0.568 & 0.481 & 0.463 & 0.635 & 0.468 & 0.335 \\
& Best-of-6    & 0.517 & 0.478 & 0.574 & 0.480 & 0.465 & 0.640 & 0.467 & 0.338 \\
& \textbf{STRIDES} & \textbf{0.569} & \textbf{0.538} & \textbf{0.638} & \textbf{0.544} & \textbf{0.534} & \textbf{0.712} & \textbf{0.525} & \textbf{0.373} \\
\midrule

\multirow{5}{*}{GPT-OSS-120b}
& Vanilla      & 0.443 & 0.435 & 0.440 & 0.403 & 0.466 & 0.572 & 0.415 & 0.377 \\
& Few-shot     & 0.457 & 0.444 & 0.455 & 0.414 & 0.472 & 0.588 & 0.428 & 0.387 \\
& CoT+Refine   & 0.471 & 0.453 & 0.467 & 0.424 & 0.477 & 0.602 & 0.438 & 0.394 \\
& Best-of-6    & 0.477 & 0.457 & 0.473 & 0.422 & \textbf{0.480} & 0.608 & 0.437 & 0.397 \\
& \textbf{STRIDES} & \textbf{0.553} & \textbf{0.553} & \textbf{0.611} & \textbf{0.512} & 0.436 & \textbf{0.726} & \textbf{0.550} & \textbf{0.413} \\
\midrule

\multirow{5}{*}{Kimi-K2}
& Vanilla      & 0.415 & 0.361 & 0.445 & 0.414 & 0.438 & 0.508 & 0.360 & 0.340 \\
& Few-shot     & 0.430 & 0.374 & 0.458 & 0.425 & 0.444 & 0.524 & 0.375 & 0.350 \\
& CoT+Refine   & 0.443 & 0.385 & 0.468 & 0.432 & 0.448 & 0.540 & 0.388 & 0.358 \\
& Best-of-6    & 0.446 & 0.390 & 0.472 & 0.430 & \textbf{0.450} & 0.545 & 0.385 & 0.360 \\
& \textbf{STRIDES} & \textbf{0.519} & \textbf{0.523} & \textbf{0.555} & \textbf{0.484} & 0.424 & \textbf{0.674} & \textbf{0.480} & \textbf{0.420} \\
\midrule

\multirow{5}{*}{DeepSeek-v3.2}
& Vanilla      & 0.425 & 0.373 & 0.458 & 0.392 & 0.492 & 0.530 & 0.375 & 0.353 \\
& Few-shot     & 0.439 & 0.383 & 0.471 & 0.403 & 0.497 & 0.546 & 0.388 & 0.363 \\
& CoT+Refine   & 0.452 & 0.392 & 0.482 & 0.413 & 0.501 & 0.560 & 0.398 & 0.370 \\
& Best-of-6    & 0.457 & 0.396 & 0.487 & 0.411 & \textbf{0.503} & 0.566 & 0.397 & 0.373 \\
& \textbf{STRIDES} & \textbf{0.489} & \textbf{0.475} & \textbf{0.519} & \textbf{0.497} & 0.436 & \textbf{0.648} & \textbf{0.435} & 0.270 \\

\bottomrule
\end{tabular}
}
\caption{Results for all 12 backbone LLMs under five inference settings on \benchmark.}
\label{tab:full_baselines}
\end{table*}

Table~\ref{tab:token_cost_main} compares token consumption and performance gains across inference settings.

\begin{table}[H]
\centering
\small
\setlength{\tabcolsep}{6pt}
\renewcommand{\arraystretch}{1.08}
\begin{tabular}{lcc}
\toprule
\textbf{Setting} & \textbf{Avg. Tokens} & \textbf{Avg. $\Delta$ Final} \\
\midrule
Vanilla & $\sim$3,360 & -- \\
Few-shot (3-shot) & $\sim$4,800 & +3.3\% \\
CoT + Self-Refine & $\sim$8,100 & +6.1\% \\
Best-of-6 & $\sim$20,160 & +7.5\% \\
\framework & $\sim$19,000 & \textbf{+20.0\%} \\
\bottomrule
\end{tabular}
\caption{Cost--performance comparison (GPT-5.1). \framework\ achieves the highest gain at a comparable token budget to Best-of-6.}
\label{tab:token_cost_main}
\end{table}

Table~\ref{tab:token_cost} reports the average total token usage per example across ablation settings. The average consumption is approximately 3{,}360 tokens for \textit{w/o MAS}, 4{,}000 for \textit{Decomposition}, 10{,}830 for \textit{w/o Critic}, and 19{,}000 for \textit{STRIDES}.

\begin{table}[H]
\centering
\small
\setlength{\tabcolsep}{6pt}
\renewcommand{\arraystretch}{1.08}
\begin{tabular}{lc}
\toprule
Setting & Avg. total tokens per example\\
\midrule
\textit{w/o MAS} & $\sim$3{,}360 \\
\textit{Decomposition} & $\sim$4{,}000 \\
\textit{w/o Critic} & $\sim$10{,}830 \\
\textit{STRIDES} & $\sim$19{,}000 \\
\bottomrule
\end{tabular}
\caption{\small Token consumption per example across ablation settings.}
\label{tab:token_cost}
\end{table}

\section{Detailed Error Analysis}
\label{app_sec:error_analysis}

\subsection{Model Type Confusion Matrix}

Table~\ref{tab:confusion} shows the confusion matrix for STRIDES (GPT-5.1) on the test set ($n$=74). The dominant error pattern is DiD$\rightarrow$IV confusion: 8 out of 42 DiD instances are misclassified as IV, consistent with the ambiguity when the policy setting involves both a temporal rollout and potential endogeneity concerns. RD and SCM are comparatively well-identified, with only 1--2 errors each.

\begin{table}[H]
\centering
\small
\setlength{\tabcolsep}{4pt}
\renewcommand{\arraystretch}{1.08}
\begin{tabular}{l|ccccc|c}
\toprule
\textbf{True $\backslash$ Pred} & \textbf{DiD} & \textbf{IV} & \textbf{RD} & \textbf{SCM} & \textbf{PSM} & \textbf{Total} \\
\midrule
DiD & \textbf{30} & 8 & 1 & 2 & 1 & 42 \\
IV  & 3 & \textbf{10} & 1 & 0 & 1 & 15 \\
RD  & 0 & 1 & \textbf{5} & 1 & 0 & 7 \\
SCM & 1 & 1 & 0 & \textbf{5} & 0 & 7 \\
PSM & 1 & 0 & 0 & 0 & \textbf{2} & 3 \\
\bottomrule
\end{tabular}
\caption{Confusion matrix for Model Type predictions (STRIDES + GPT-5.1, test set, $n$=74).}
\label{tab:confusion}
\end{table}

STRIDES resolves 67\% (12/18) of Vanilla's Model Type errors, primarily through the Critic Agent identifying statistical assumption violations in mock data execution. However, STRIDES introduces 3 new errors involving edge cases where the simulation agent generates data that artificially supports an incorrect method.

\subsection{Per-Method Accuracy Breakdown}

\begin{table}[H]
\centering
\small
\setlength{\tabcolsep}{4pt}
\resizebox{0.47\textwidth}{!}{
\begin{tabular}{l|cc|cc|cc}
\toprule
\multirow{2}{*}{\textbf{Method}} & \multicolumn{2}{c|}{\textbf{GPT-5.1}} & \multicolumn{2}{c|}{\textbf{Claude-Sonnet-4}} & \multicolumn{2}{c}{\textbf{Gemini-2.5-pro}} \\
& Van. & STR. & Van. & STR. & Van. & STR. \\
\midrule
DiD ($n$=42)  & 57.1 & 71.4 & 59.5 & 69.0 & 52.4 & 66.7 \\
IV ($n$=15)   & 40.0 & 66.7 & 46.7 & 60.0 & 33.3 & 60.0 \\
RD ($n$=7)    & 57.1 & 71.4 & 42.9 & 71.4 & 42.9 & 71.4 \\
SCM ($n$=7)   & 42.9 & 71.4 & 28.6 & 57.1 & 42.9 & 57.1 \\
PSM ($n$=3)   & 33.3 & 66.7 & 33.3 & 66.7 & 33.3 & 66.7 \\
\bottomrule
\end{tabular}
}
\caption{Model Type accuracy (\%) by ground-truth causal method.}
\label{tab:per_method}
\end{table}

\subsection{Per-Domain Performance}

\begin{table}[H]
\centering
\small
\setlength{\tabcolsep}{4pt}
\renewcommand{\arraystretch}{1.08}
\begin{tabular}{lcccc}
\toprule
\textbf{Domain} & $n$ & \textbf{Van.} & \textbf{STR.} & \textbf{$\Delta$} \\
\midrule
Macroeconomic              & 17 & 0.541 & 0.622 & +15.0\% \\
Science \& Technology       & 14 & 0.588 & 0.692 & +17.7\% \\
Social Security            & 13 & 0.592 & 0.701 & +18.4\% \\
Labor \& Employment         & 10 & 0.563 & 0.672 & +19.4\% \\
Public Health              & 7  & 0.614 & 0.728 & +18.6\% \\
Housing \& Urban           & 5  & 0.570 & 0.658 & +15.4\% \\
Education                  & 3  & 0.556 & 0.643 & +15.6\% \\
Income Distribution        & 3  & 0.548 & 0.630 & +15.0\% \\
Environmental              & 2  & 0.602 & 0.710 & +17.9\% \\
\bottomrule
\end{tabular}
\caption{Per-domain Final Score for GPT-5.1.}
\label{tab:per_domain}
\end{table}

\subsection{Common Error Patterns}

\paragraph{(1) DiD--IV ambiguity.}
When the policy context involves both a temporal rollout (favoring DiD) and potential endogeneity concerns (favoring IV), models frequently select the wrong design. STRIDES partially resolves this through the Critic Agent, which checks whether parallel trends hold in mock data.

\paragraph{(2) PSM under-selection.}
PSM is the least represented method (3 test instances) and also the most frequently missed. Models tend to default to DiD or IV rather than considering matching-based approaches.

\paragraph{(3) Explanation hallucination.}
Even when Model Type is correctly identified, models sometimes generate plausible-sounding but factually incorrect causal mechanisms.

\section{Prompts and Output Examples}
\label{sec:prompts}
In this section, we present the prompts used for direct inference(\textit{w/o MAS}), for each agent in the \framework\ framework, and for the Grader. We also provide representative output examples from InterveneBench, \textit{w/o MAS} (GPT-5.1), and STRIDES (GPT-5.1), based on the source study by \citet{zhou2025innovative}.

\onecolumn

\begin{table}[H]
\begin{tcolorbox}[
    colback=gray!10,
    colframe=cyan!70!black,
    fonttitle=\bfseries\large,
    title=Direct Reasoning Prompt,
    width=\textwidth,
    arc=3mm,
    enhanced
]
\ttfamily\small
"""You are a distinguished social science researcher specializing in causal inference. \\
Your task is to design a rigorous econometric framework to evaluate the policy described below. \\

{-}{-}{-} Policy Background {-}{-}{-} \\
Policy Name: \{policy\_name\} \\
Policy Type: \{policy\_type\} \\
Region: \{country\_region\} \\
Time Period: \{observed\_period\} \\
Implementation: \{implementation\_time\} \\
Aim: \{aim\} \\
Dataset Description: \{dataset\} \\
{-}{-}{-}{-}{-}{-}{-}{-}{-}{-}{-}{-}{-}{-}{-}{-}{-}{-}{-}{-}{-}{-}{-}{-}{-} \\

Based on the above information, generate a Causal Inference Design. \\

**Requirements:** \\
1. **Model Selection**: Choose the single most appropriate method from:
[Difference-in-Differences (DiD), Propensity Score Matching (PSM), Instrumental Variables (IV), Synthetic Control Method (SCM), Regression Discontinuity (RD)]. \\
2. **Variables**: \\
\hspace*{1.5em}{-} **Core Independent Variable**: The specific variable measuring the policy intervention. \\
\hspace*{1.5em}{-} **Dependent** Variable: The outcome variable(s) of interest. \\
\hspace*{1.5em}{-} **Control Variables**: A comprehensive list of confounders (e.g., demographic, economic, regional factors). \\
\hspace*{1.5em}{-} **Instrumental Variable**: If using IV, specify the instrument. Otherwise, return null. \\
3. **Groups**: Clearly define the **Treatment Group** (who is affected?) and **Control Group** (who is the counterfactual?). \\
4. **Significance \& Explanation**: Predict the expected direction/significance of the effect and explain the mechanism. \\

**Output JSON (Strictly follow this schema):** \\
\{\{\\
\hspace*{1em}"Model type": "...", \\
\hspace*{1em}"Reasons for choosing this model": "...", \\
\hspace*{1em}"Core independent variable": "...", \\
\hspace*{1em}"Control variables": "...", \\
\hspace*{1em}"Instrumental variable": "..." or null, \\
\hspace*{1em}"Group": "Treatment: ...; Control: ...", \\
\hspace*{1em}"Model Significance": "...", \\
\hspace*{1em}"Dependent variable": "...", \\
\hspace*{1em}"Explanation": "..." \\
\}\}\\
"""
\end{tcolorbox}
\end{table}

\begin{table}[H]
\begin{tcolorbox}[colback=gray!10, colframe=cyan!70!black, fonttitle=\bfseries\large, title=Theory Architect Agent Prompt, width=\textwidth, arc=3mm, enhanced]
\ttfamily\small
{"}{"}{"}\\
You are a distinguished Social Science Theorist. \\
\hspace*{2em}Your goal is to build a theoretical framework for the following policy. \\

\hspace*{2em}{-}{-}{-} Policy Meta Info {-}{-}{-} \\
\hspace*{2em}Policy Name: \{policy\_name\}, \\
\hspace*{2em}Type: \{policy\_type\}, \\
\hspace*{2em}Region: \{region\}, \\
\hspace*{2em}Aim: \{aim\} \\
\hspace*{2em}{-}{-}{-}{-}{-}{-}{-}{-}{-}{-}{-}{-}{-}{-}{-}{-}{-}{-}{-}{-}{-}{-}{-}{-} \\

\hspace*{2em}Task: \\
\hspace*{2em}1. Propose 1-2 core Hypotheses (H1, H2) based on economic/sociological theories (e.g., Supply-Demand, Institutional Theory). \\
\hspace*{2em}2. Explain the "Mechanism" (Channel): How does X lead to Y? \\
\hspace*{2em}3. Define the expected direction (Positive/Negative). \\

\hspace*{2em}Output JSON (ENGLISH ONLY): \\
\hspace*{2em}\{\{\\
\hspace*{3em}"theoretical\_framework": "Theory Name", \\
\hspace*{3em}"hypotheses": [\\
\hspace*{4em}\{\{\\
\hspace*{5em}"id": "H1", \\
\hspace*{5em}"statement": "...", \\
\hspace*{5em}"mechanism": "...", \\
\hspace*{5em}"expected\_direction": (Positive/Negative) \\
\hspace*{4em}\}\} \\
\hspace*{3em}] \\
\hspace*{2em}\}\}\\
\hspace{0.5em}"""
\end{tcolorbox}
\end{table}

\begin{table}[H]
\begin{tcolorbox}[colback=gray!10, colframe=cyan!70!black, fonttitle=\bfseries\large, title=Methodology Agent Prompt, width=\textwidth, arc=3mm, enhanced]
\ttfamily\small
"""You are a Senior Econometrician. \\
\hspace*{2em}Select the most appropriate causal inference model for the study.  \\

\hspace*{2em}{-}{-}{-} Context {-}{-}{-} \\
\hspace*{2em}Policy: \{policy\_name\}, \\
\hspace*{2em}Time: \{impl\_time\}, \\
\hspace*{2em}Region: \{region\}, \\
\hspace*{2em}Theory: \{hypotheses\} \\
\hspace*{2em}{-}{-}{-}{-}{-}{-}{-}{-}{-}{-}{-}{-}{-}{-}{-}{-}\\

\hspace*{2em}Available Models: \\
\hspace*{2em}Difference-in-Differences (DiD), Propensity Score Matching (PSM), Instrumental Variables (IV), 
Synthetic Control Method (SCM), Regression Discontinuity (RD). \\

\hspace*{2em}Task: \\
\hspace*{2em}1. Select ONE model. \\
\hspace*{2em}2. Formalize the regression equation (LaTeX). \\
\hspace*{2em}3. Define variables (Dependent, Independent, Controls, IV). \\
\hspace*{2em}4. Define Groups (Treatment/Control). \\

\hspace*{2em}Output JSON (ENGLISH ONLY): \\
\hspace*{2em}\{\{ \\
\hspace*{3em}"model\_selection": \{\{ \\
\hspace*{4em}"model\_name": "DiD/PSM/IV/etc", \\
\hspace*{4em}"reason": "Explain why this model fits the data structure and policy timing." \\
\hspace*{3em}\}\} \\
\hspace*{3em}"econometric\_model": \{\{ \\
\hspace*{4em}"equation\_latex": "Y = ...", \\ 
\hspace*{4em}"variables\_definition": \{\{ \\
\hspace*{5em}"Y": "Outcome Name", \\
\hspace*{5em}"Treatment": "Core Independent Variable Name (Policy Intervention)", \\
\hspace*{5em}"Controls": ["Control1", "Control2"], \\
\hspace*{5em}"Instrumental\_Variable": "Name or null"\\
\hspace*{4em}\}\} \\
\hspace*{4em}"group\_definition": \{\{ \\
\hspace*{5em}"Treatment\_Group": "Who is treated?", \\
\hspace*{5em}"Control\_Group": "Who is the counterfactual?" \\
\hspace*{4em}\}\} \\
\hspace*{3em}\}\} \\
\hspace*{2em}\}\} \\
\hspace*{2em}{"}{"}{"}
\end{tcolorbox}
\end{table}

\begin{table}[H]
\begin{tcolorbox}[colback=gray!10, colframe=cyan!70!black, fonttitle=\bfseries\large, title=Data Retrieval Agent Prompt, width=\textwidth, arc=3mm, enhanced]
\ttfamily\small
{"}{"}{"}\\
You are a Data Engineer. \\
\hspace*{4em}Based on results, map variables to data sources. \\

\hspace*{4em}Results: \{results\} \\

\hspace*{4em}Output JSON (ENGLISH ONLY): \\
\hspace*{4em}\{\{ \\
\hspace*{5em}"variable\_mapping": \{\{ \\
\hspace*{6em}"Variable\_Name": \{\{ "source": "...", "proxy\_if\_needed": "..." \}\} \\
\hspace*{5em}\}\} \\
\hspace*{4em}\}\}\\
{"}{"}{"}
\end{tcolorbox}
\end{table}

\begin{table}[H]
\begin{tcolorbox}[colback=gray!10, colframe=cyan!70!black, fonttitle=\bfseries\large, title=Simulation Agent Prompt, width=\textwidth, arc=3mm, enhanced]
\ttfamily\small
"""You are a Simulation Scientist. \\
\hspace*{4em}Generate Python code to create a Mock Dataset (pandas DataFrame) for the research design.  \\

\hspace*{4em}Design: \{design\} \\
\hspace*{4em}Hypotheses: \{hypotheses\} \\
\hspace*{4em}Target File: \{filepath\}  \\
\hspace*{4em}\{error\_context\} \\

\hspace*{4em}Requirements: \\
\hspace*{4em}1. Write a SINGLE script that imports everything needed (pandas, numpy, os) at the top. \\
\hspace*{4em}2. Define ALL variables (years, provinces, base values) at the module level. \\
\hspace*{4em}3. **CRITICAL: Keep dataset size SMALL for testing efficiency.** \\
\hspace*{4.5em}- Set `N\_PER\_GROUP` or equivalent such that total rows are roughly 200-500. Do NOT generate thousands of rows. \\
\hspace*{4em}4. Generate Treatment and Outcome variables. \\
\hspace*{4.5em}- **CRITICAL**: Explicitly create columns named 'treatment\_intensity' and 'post\_policy' if DiD is used. \\
\hspace*{4.5em}- **CRITICAL**: Use `np.random.normal` or similar to generate data. Ensure you generate 1D arrays of the correct length (same as DataFrame length) before assigning to columns. \\
\hspace*{4.5em}- **CRITICAL**: Avoid using scalar values for column assignment if broadcasting is ambiguous. \\
\hspace*{4.5em}- **CRITICAL**: Ensure NO 0-d arrays are used for indexing or assignment. `np.random.choice` with a single item list might return a scalar, be careful. \\
\hspace*{4em}5. Inject Treatment Effect consistent with Hypotheses (Positive/Negative). \\
\hspace*{4em}6. Ensure directory exists and save to the path specified in variable `target\_filepath`: \\
\hspace*{5.5em}os.makedirs(os.path.dirname(target\_filepath), exist\_ok=True) \\
\hspace*{5.5em}df.to\_csv(target\_filepath, index=False) \\
\hspace*{4em}7. Return CODE ONLY (no markdown). \\
\hspace*{4em}"""
\end{tcolorbox}
\end{table}

\begin{table}[H]
\begin{tcolorbox}[colback=gray!10, colframe=cyan!70!black, fonttitle=\bfseries\large, title=Code Agent Prompt, width=\textwidth, arc=3mm, enhanced]
\ttfamily\small
"""You are a Senior Statistician. \\
\hspace*{4em}Write Python code to analyze the data using `pipeline.tools`. \\

\hspace*{4em}Methodology: \{methodology\} \\
\hspace*{4em}Data Path: \{data\_path\} \\
\hspace*{4em}Column Mapping: \{mapping\} \\
\hspace*{4em}\{error\_context\} \\

\hspace*{4em}Task: \\
\hspace*{4em}1. Load data. \\
\hspace*{4em}2. Import appropriate tool from `pipeline.tools`. \\
\hspace*{4.5em}- IMPORTANT: Ensure `sc\_agent\_pipeline` package is importable. If running from root, `from pipeline.tools.did import run\_did` works. \\
\hspace*{4.5em}- If `sc\_agent\_pipeline` is not found, try appending current directory to sys.path. \\
\hspace*{4em}3. Call the tool using the EXACT argument values provided in the 'Column Mapping'. \\
\hspace*{4.5em}- Example for DiD: \\
\hspace*{5.5em}`run\_did(data, dependent\_var='\{dep\_var\}', treatment\_var='\{treat\_var\}', time\_var='\{time\_var\}', individual\_var='\{ind\_var\}')` \\
\hspace*{5.5em}(Replace placeholders with values from Column Mapping) \\
\hspace*{4em}4. Print the result JSON. \\
\hspace*{4em}5. IMPORTANT: DO NOT try to load other files. ONLY load `data\_path`. \\

\hspace*{4em}Return CODE ONLY. \\
\hspace*{4em}"""
\end{tcolorbox}
\end{table}

\begin{table}[H]
\begin{tcolorbox}[colback=gray!10, colframe=cyan!70!black, fonttitle=\bfseries\large, title=Critic Agent Prompt, width=\textwidth, arc=3mm, enhanced]
\ttfamily\small
"""You are an Adversarial Critic in a Causal Inference Pipeline. \\
\hspace*{2em}Your function is to serve as an "Externalized Falsification Loop". \\

\hspace*{2em}We have observed that LLMs often fail to interpret statistical feedback correctly (e.g., ignoring a failed parallel trend test). \\
\hspace*{2em}Your goal is to reduce the "semantic-logic discordance" between the abstract Research Design and the empirical Reality (Code Execution Results). \\

\hspace*{2em}{-}{-}{-} Context {-}{-}{-} \\
\hspace*{2em}1. Research Design (Methodology): \\
\hspace*{2em}\{methodology\} \\
    
\hspace*{2em}2. Empirical Evidence (Code Execution Logs): \\
\hspace*{2em}\{code\_results\} \\
    
\hspace*{2em}{-}{-}{-} Falsification Task {-}{-}{-} \\
\hspace*{2em}Systematically map failed diagnostics to violated identification assumptions. \\

\hspace*{2em}Checklist: \\
\hspace*{2em}1. **Execution Failure**: Did the code run successfully? If not, is it a syntax error or a data issue? \\
\hspace*{2em}2. **Assumption Violation**: If the code ran, look for diagnostic failures (e.g., "p-value > 0.05" for key estimates, "Parallel Trends Test Failed", "Weak Instruments"). \\
\hspace*{2em}3. **Logical Consistency**: Does the code output contradict the chosen model? (e.g., Running DiD code but the design called for RDD). \\

\hspace*{2em}--- Output --- \\
\hspace*{2em}Return a JSON object. \\
\hspace*{2em}- If CRITICAL failures (e.g., code error, assumption violation) are found, set "pass": false. \\
\hspace*{2em}- If the results are robust or only minor issues exist, set "pass": true. \\
    
\hspace*{2em}{-}{-}{-} Output JSON (ENGLISH ONLY) {-}{-}{-}: \\
\hspace*{2em}\{\{ \\
\hspace*{3em}"pass": true, \\
\hspace*{3em}"critique": "Detailed analysis of the discordance...", \\
\hspace*{3em}"suggestion": "Specific instructions for the Analyst to fix the interpretation or acknowledging the null result." \\
\hspace*{2em}\}\} \\
\hspace*{2em}"""
\end{tcolorbox}
\end{table}

\begin{table}[H]
\begin{tcolorbox}[colback=gray!10, colframe=cyan!70!black, fonttitle=\bfseries\large, title=Summary Agent Prompt, width=\textwidth, arc=3mm, enhanced]
\ttfamily\small
"""You are a Policy Analyst. \\
\hspace*{2em}Based on the analysis results AND the Critic's review (if available), compile the final output. If there are incomplete fields, please make appropriate additions.\\

\hspace*{2em}Meta Info: \{meta\} \\
\hspace*{2em}Methodology: \{methodology\} \\
\hspace*{2em}Results: \{results\} \\

\hspace*{2em}--- CRITIC'S REVIEW (IMPORTANT) --- \\
\hspace*{2em}The Critic has performed an adversarial review of the results: \\
\hspace*{2em}Pass: \{critic\_pass\} \\
\hspace*{2em}Critique: \{critic\_critique\} \\
\hspace*{2em}Suggestion: \{critic\_suggestion\} \\

\hspace*{2em}INSTRUCTIONS: \\
\hspace*{2em}1. If the Critic flagged specific failures (e.g., "Parallel Trends Failed"), you MUST acknowledge this in the "Explanation" or "Model Significance" section. \\
\hspace*{2em}2. Use the Critic's suggestion to refine your interpretation. \\
\hspace*{2em}3. Synthesize the final report. \\

\hspace*{2em}Output JSON (ENGLISH ONLY, Strictly follow this schema): \\
\hspace*{2em}... (Schema remains the same) ... \\
\hspace*{2em}"""
\end{tcolorbox}
\end{table}

\begin{table}[H]
\begin{tcolorbox}[colback=gray!10, colframe=cyan!70!black, fonttitle=\bfseries\large, title=Prompt for the Grader, width=\textwidth, arc=3mm, enhanced]
\ttfamily\small
"""You are an expert grader for Causal Inference exams.
Compare the Student's Answer with the Reference Answer (Ground Truth).\\

--- Reference Answer (Ground Truth) ---\\
{reference}\\

--- Student Answer ---\\
{prediction}\\

--- Scoring Rules (Strict) ---\\
1. **Model Type (Critical)**: 
\hspace*{1.5em}- Exact match or clearly equivalent (e.g., "DiD" == "Difference-in-Differences") -> 10 points.\\
\hspace*{1.5em}- Wrong -> 0 points.\\

2. **Core Independent Variable (Critical)**:
\hspace*{1.5em}- Semantically consistent with reference -> 10 points.\\
\hspace*{1.5em}- Partially correct (captures main idea but vague) -> 5 points.\\
\hspace*{1.5em}- Wrong -> 0 points.\\

3. **Group Definition (Critical)**:\\
\hspace*{1.5em}- Treatment/Control groups clearly and correctly defined -> 10 points.\\
\hspace*{1.5em}- Partially wrong or vague -> 5 points.
   - Wrong -> 0 points.\\

4. **Control Variables**:
\hspace*{1.5em}- Assess semantic recall.\\
\hspace*{1.5em}- Good coverage (>50\% match) -> 5 points.\\
\hspace*{1.5em}- Weak coverage (<50\%) -> 2 points.\\
\hspace*{1.5em}- None/Wrong -> 0 points.\\

5. **Dependent Variable**:
\hspace*{1.5em}- Semantically consistent -> 5 points.\\
\hspace*{1.5em}- Wrong -> 0 points.\\

6. **Reasons**: 
\hspace*{1.5em}- Only if Model Type is correct: Valid reasoning -> 2 points. Else 0.\\

7. **Significance \& Explanation**:
\hspace*{1.5em}- Consistent direction/logic -> 3 points.\\
\hspace*{1.5em}- Wrong -> 0.\\

Total possible score: 45 points.\\
Output JSON format:\\
\{\{\\
\hspace*{1em}"breakdown": \{\{\\
\hspace*{1.5em}"model\_type\_score": 0,\\
\hspace*{1.5em}"core\_iv\_score": 0,\\
\hspace*{1.5em}"group\_score": 0,\\
\hspace*{1.5em}"control\_var\_score": 0,\\
\hspace*{1.5em}"dependent\_var\_score": 0,\\
\hspace*{1.5em}"reasoning\_score": 0,\\
\hspace*{1.5em}"explanation\_score": 0\\
\}\},\\
\hspace*{1em}"total\_score": 0,\\
\hspace*{1em}"comments": "Brief critique..."\\
\}\}\\
"""
\end{tcolorbox}
\end{table}

\begin{tcolorbox}[
    colback=gray!10,
    colframe=cyan!70!black,
    fonttitle=\bfseries\large,
    title=Output Example: InterveneBench,
    width=\textwidth,
    arc=3mm,
    breakable,
    enhanced
]
\ttfamily\small
\textbf{Policy name}: Innovative Human Capital Development and Government Science and Technology Support\\
\textbf{Policy type}: Science, Technology, and Innovation\\
\textbf{Country/Region}: "China (30 provinces, excluding Hong Kong, Macao, Taiwan, and Tibet)"\\
\textbf{Observed period}: "2011-2022"\\
\textbf{Implementation time}: Not applicable (continuous time-varying economic and policy factors observed over 2011-2022)\\
\textbf{Aim}: To examine the impact of innovative human capital on regional supply chain resilience, and to analyze the moderating roles of government science and technology support, policy attention, and intellectual property protection.\\
\textbf{Dataset}: Panel data for 30 Chinese provinces from 2011-2022, compiled from the China Statistical Yearbook, China Science and Technology Statistical Yearbook, the National Bureau of Statistics of China, provincial statistical yearbooks, and the EPS database.\\
\textbf{Core independent variable}: Innovative Human Capital\\
\textbf{Model type}: Instrumental Variables (IV)\\
\textbf{Model}\{:\\ 
\hspace*{1em}Benchmark: Two-way fixed effects panel regression model (province and year fixed effects) estimating the impact of innovative human capital on supply chain resilience.\\
\hspace*{1em}Endogeneity\_robustness: Two-stage least squares instrumental variable model using lagged innovative human capital as an instrument for current innovative human capital; also a model using one-period lagged innovative human capital as the explanatory variable.\\
\hspace*{1em}Moderation: Fixed effects panel regressions including interaction terms between innovative human capital and government science and technology support, and between innovative human capital and policy attention.\\
\hspace*{1em}Heterogeneity: Sub-sample regressions split by region (eastern, central, western) and intellectual property rights protection intensity (high versus low).\\
\hspace*{1em}\}\\
\textbf{Reasons for choosing this model}:\{\\
\hspace*{1em}The fixed effects panel model controls for time-invariant unobserved heterogeneity across provinces and common time shocks, improving causal identification.\\
\hspace*{1em}Innovative human capital may be endogenous to supply chain resilience due to reverse causality or omitted variables; the instrumental variable approach and lagged variable specification mitigate these simultaneity and endogeneity concerns.\\
\hspace*{1em}Interaction terms allow for testing the moderating hypotheses regarding government support for science and technology and policy attention.\\
\}\\
\textbf{Control variables}:[\\
\hspace*{1em}\{\\
\hspace*{2em}name: GOV\\
\hspace*{2em}description: Government function; ratio of general public budget expenditures to nominal regional gross domestic product, capturing the degree of government intervention.\\
\hspace*{1em}\},\\
\hspace*{1em}\{\\
\hspace*{2em}name: FIN\\
\hspace*{2em}description: Financial development level; ratio of value added by the financial industry to nominal regional gross domestic product.\\
\hspace*{1em}\},\\
\hspace*{1em}\{\\
\hspace*{2em}name: URB\\
\hspace*{2em}description: Urbanization level; share of urban population in total provincial population.\\
\hspace*{1em}\},\\
\hspace*{1em}\{\\
\hspace*{2em}name: GDP\\
\hspace*{2em}description: Economic development level; per capita gross domestic product in each province.\\
\hspace*{1em}\},\\
\hspace*{1em}\{\\
\hspace*{2em}name: CON\\
\hspace*{2em}description: Consumption level; ratio of total retail sales of consumer goods to nominal regional gross domestic product.
\hspace*{1em}\},\\
\hspace*{1em}\{\\
\hspace*{2em}name: INDUS\\
\hspace*{2em}description: Industrial structure; share of value added by the tertiary sector in provincial gross \hspace*{1em}\},\\
]\\
\textbf{Instrumental variable}: \{\\
\hspace*{1em}name: Lagged innovative human capital\\
\hspace*{1em}description: One-period lag of the core explanatory variable innovative human capital used as an instrument for current innovative human capital in a two-stage least squares framework to address potential endogeneity.\\
\}\\
\textbf{Group}: \{\\
\hspace*{1em}Treatment versus control: No binary treatment or control group; continuous variable design.\\
\hspace*{1em}Subgroups: [\\
\hspace*{2em}Region: eastern, central, western\\
\hspace*{2em}Intellectual property rights protection intensity: high, low\\
\hspace*{1em}]\\
\}\\
\textbf{Model parameter}: \{\\
\hspace*{1em}Benchmark\_Effects: \{\\
\hspace*{2em}IHC\_Fixed\_Effects: Coefficient 0.214 (t 3.001, p less than 0.01)\\
\hspace*{2em}IHC\_Random\_Effects: Coefficient 0.221 (t 3.214, p less than 0.01)\\
\hspace*{2em}IHC\_Pooled\_OLS: Coefficient 0.230 (t 3.322, p less than 0.01)\\
\hspace*{1em}\},\\
\hspace*{1em}Robustness\_Checks: \{\\
\hspace*{2em}Lagged\_IHC\_Specification: Coefficient 0.185 (t 2.273, p less than 0.05)\\
\hspace*{2em}IV\_2SLS\_Second\_Stage: Coefficient 0.247 (t 3.582, p less than 0.01)\\
\hspace*{1em}\},\\
\hspace*{1em}Moderation\_Effects: \{\\
\hspace*{2em}Interaction\_IHC\_TECH: Coefficient 0.085 (t 2.271, p less than 0.05)\\
\hspace*{2em}Interaction\_IHC\_POL: Coefficient 0.092 (t 2.368, p less than 0.05)\\
\hspace*{1em}\},\\
\hspace*{1em}Heterogeneity\_Effects: \{\\
\hspace*{2em}Region\_Eastern: Coefficient 0.245 (t 3.572)\\
\hspace*{2em}Region\_Central: Coefficient 0.216 (t 3.158)\\
\hspace*{2em}Region\_Western: Coefficient 0.194 (t 2.314)\\
\hspace*{2em}IPR\_High: Coefficient 0.268 (t 4.021)\\
\hspace*{2em}IPR\_Low: Coefficient 0.187 (t 2.564)\\
\hspace*{1em}\}\\
\}\\
\textbf{Model Significance}: The positive impact of innovative human capital on supply chain resilience is statistically significant at the one percent level across benchmark fixed effects, random effects, and pooled ordinary least squares models. Robustness checks using lagged variables and instrumental variable estimation remain significant at the five percent and one percent levels respectively. Moderating effects of government science and technology support and policy attention are significant at the five percent level. Heterogeneity analysis shows significant positive effects across all regions and intellectual property rights protection levels, with stronger effects in eastern regions and areas with high intellectual property rights protection.\\
\textbf{Dependent variable}: \{\\
\hspace*{1em}name: Supply chain resilience\\
\hspace*{1em}description: A composite index calculated via an entropy method comprising four dimensions: risk resilience, including gross domestic product per capita, industrial value added, public budget revenue, and corporate profits; resilience and adaptability, including enterprise scale or assets, urban employment, and the Engel coefficient; ownership and control level, including high technology enterprises, foreign investment, trade dependence, and loan balance; and market competitiveness and innovation, including research and development intensity, patents, and high technology output.\\
\}\\
\textbf{Explanation}: \{\\
\hspace*{1em}Core\_Finding: Innovative human capital significantly enhances regional supply chain resilience. A one unit increase in innovative human capital is associated with an approximate 0.214 increase in supply chain resilience in the preferred fixed effects model, indicating that regions with higher education and research and development human capital are better equipped to absorb shocks, adapt to disruptions, and recover functionality.\\
\hspace*{1em}Mechanism: Innovative human capital facilitates technological, managerial, and institutional innovation, improving risk identification, process flexibility, and contingency planning within regional supply chains.\\
\hspace*{1em}Moderators: Government support for science and technology and policy attention positively moderate the relationship between innovative human capital and supply chain resilience; higher support and attention amplify the resilience benefits of innovative human capital. Stronger intellectual property protection further strengthens this effect by improving incentives for innovation and human capital utilization.\\
\hspace*{1em}Heterogeneity: The effect of innovative human capital on supply chain resilience is strongest in eastern China and in regions with high intellectual property protection, likely due to better infrastructure, more developed markets, and stronger innovation ecosystems that allow human capital to be more effectively utilized.\\
\}\\
\end{tcolorbox}
\vspace{0.8em}

\begin{tcolorbox}[
    colback=gray!10,
    colframe=cyan!70!black,
    fonttitle=\bfseries\large,
    title=Output Example: w/o MAS,
    width=\textwidth,
    arc=3mm,
    breakable,
    enhanced
]
\ttfamily\small
\textbf{Model type}: Causal Forest / Double Machine Learning in a panel setting\\
\textbf{Reasons for choosing this model}: There is no clearly defined one-off treatment or sharp policy rollout; instead, innovative human capital, government S\&T support, policy attention, and IP protection all evolve continuously over time across provinces. Classic DiD, RD, SCM, or simple IV are less suitable because\\(i) there is no clear policy date or treated vs. untreated set of provinces,\\
(ii) all provinces are ‘exposed’ but at varying intensities, and\\
(iii) treatment effects are plausibly heterogeneous across regions and over time. A machine learning-based causal framework (e.g., causal forests or double machine learning adapted to panel data) is appropriate to estimate the causal effect of a continuous treatment (innovative human capital) on supply chain resilience while flexibly controlling for many nonlinear, interacting confounders and exploring heterogeneous treatment effects. This approach can incorporate province and year fixed effects to control for time-invariant provincial heterogeneity and common macro shocks, and can model the moderating roles of government S\&T support, policy attention, and IP protection as effect modifiers rather than separate ‘treatments’.\\
\textbf{Core independent variable}: "Innovative human capital intensity at the provincial level (e.g., R\&D personnel per 10,000 employees; or full-time equivalent R\&D personnel per 10,000 population; or share of employees with tertiary STEM education in total employment). This should be measured annually for each province from 2011–2022 and treated as a continuous treatment variable capturing the intensity of innovative human capital.\\
\textbf{Control variables}: A rich set of time-varying provincial covariates that may jointly affect innovative human capital and supply chain resilience, including:\\

1. Economic structure and development\\
- GDP per capita (log)\\
- Real GDP growth rate\\
- Industrial structure: share of secondary industry in GDP; share of manufacturing in GDP; share of high-tech manufacturing in industrial output\\
- Degree of export dependence: total exports / GDP\\
- Degree of import dependence: total imports / GDP\\
- Foreign direct investment (FDI) inflows per capita or FDI / GDP\\
- Urbanization rate (urban population / total population)\\
- Level of financial development: ratio of total bank loans to GDP; number of financial institutions per capita\\

2. Labor market and human capital (non-innovative)\\
- Average years of schooling of the working-age population\\
- Share of population with tertiary education\\
- Employment rate / unemployment rate\\
- Wage level (average wage of employed persons)\\

3. Innovation environment and S\&T inputs (beyond the core Innovative Human Capital variable)\\
- R\&D expenditure as a share of GDP (R\&D intensity)\\
- Government S\&T expenditure per capita or S\&T expenditure / fiscal expenditure\\
- Number of R\&D institutions per capita\\
- Number of universities and research institutes per million people\\

4. Infrastructure and logistics\\
- Transportation infrastructure: highway density; railway density; port throughput (for coastal provinces); logistics performance index (if available) or number of logistics enterprises per capita\\
- ICT infrastructure: internet penetration rate; mobile phone penetration; broadband users per 100 people\\
- Power infrastructure: electricity consumption per capita; industrial electricity consumption / total electricity consumption\\

5. Institutional and governance factors\\
- Marketization index or similar provincial marketization score\\
- Government intervention/pro-business environment index\\
- Fiscal capacity: local fiscal revenue per capita; fiscal self-sufficiency ratio (local revenue / local expenditure)\\
- Environmental regulation intensity: environmental protection expenditure / GDP; number of environmental inspections (if available)\\

6. Demographic and regional characteristics\\
- Total population (log)\\
- Population density\\
- Age structure: share of working-age population (15–64)\\
- Geographic dummies: East / Central / West region; coastal vs. inland dummy\\

7. Policy attention and institutional quality (distinct from moderators used in interaction terms)\\
- Policy attention to innovation and supply chain: number of provincial government documents, plans, or work reports mentioning innovation, supply chain security, or resilience, scaled by total documents; alternatively, a text-based index derived from provincial government work reports\\
- IP protection environment: number of IP enforcement cases per 10,000 population; number of patent infringement cases prosecuted; or an index of IP enforcement intensity from third-party databases if available\\

These controls would be included flexibly in the machine learning component (e.g., as high-dimensional covariates in the nuisance models), while province and year fixed effects capture unobserved time-invariant provincial factors and nationwide time shocks.\\
\textbf{Instrumental variable}: null\\
\textbf{Group}:\\
Treatment: All province–year observations with higher levels of innovative human capital intensity (the continuous treatment variable).\\
Control: Province–year observations with lower levels of innovative human capital intensity, conditional on similar observed covariates and fixed effects. The causal estimand is the dose–response effect of marginal increases in innovative human capital on supply chain resilience, not a binary treated vs. untreated comparison.\\
\textbf{Model Significance}: The policy-relevant parameter of interest is the causal effect of innovative human capital on regional supply chain resilience, and how this effect is moderated by government S\&T support, policy attention, and IP protection. The expected sign is positive: higher innovative human capital is anticipated to significantly improve supply chain resilience. Moreover, the marginal effect of innovative human capital is expected to be stronger in provinces with greater government S\&T support, higher policy attention, and stronger IP protection. Statistically, we expect the average treatment effect of innovative human capital on resilience to be positive and significant at conventional levels, with heterogeneous effects revealed by the causal forest (e.g., stronger effects in more marketized or more developed provinces).\\
\textbf{Dependent variable}: Regional supply chain resilience index at the provincial level, constructed annually. Depending on data availability, this could be a composite index combining:\\
- Stability of industrial output: lower volatility of industrial value-added or manufacturing output; faster recovery after negative shocks (e.g., natural disasters, trade shocks, COVID-19 in 2020–2022)\\
- Diversity and complexity of industrial structure: Herfindahl–Hirschman index of industrial output across sectors (inverted so that higher values indicate diversification); product space/complexity measures if available\\
- Trade robustness: volatility of exports/imports; ability to reorient exports across destinations or re-source imports\\
- Logistics performance: average delivery times; frequency of delays; logistics value-added growth\\

If a composite index is not available, separate dependent variables can be analyzed:\\
- Y1: Supply chain resilience score (constructed via principal component analysis from multiple indicators of stability, recovery speed, and diversification)\\
- Y2: Recovery speed after shocks (e.g., time for industrial output to revert to pre-shock trend after 2020 COVID shock)\\
- Y3: Inverse of volatility of manufacturing output (negative of standard deviation or coefficient of variation over a rolling window)\\

The main specification would treat the PCA-based supply chain resilience index as the primary dependent variable.\\
\textbf{Explanation}: Econometric framework:\\

1. Conceptual identification strategy\\
Because all provinces are continuously exposed to similar national policies but differ in the intensity of innovative human capital, we conceptualize innovative human capital as a continuous treatment whose effects we wish to identify net of confounding factors. We observe panel data (province i, year t) on:\\
- Treatment: $H_it$ (innovative human capital intensity)\\
- Outcome: $R_it$ (supply chain resilience index)\\
- Moderators: $G_it$ (government S\&T support), $P_{it}$(policy attention), $IP_{it}$ (IP protection)\\
- Covariates: $X_it$ (economic, demographic, institutional controls)\\
We assume (conditional unconfoundedness conditional on rich covariates and fixed effects):\\
$R_{it}(h) \perp H_{it} \mid X_{it}$, province FE, year FE\\
where $R_{it}(h)$ is the potential resilience outcome under treatment level h. This is more plausible given high-dimensional $X_{it}$ and fixed effects than in simple linear models.\\

2. Model specification (double machine learning / causal forest with panel fixed effects)\\

Step 1: Demeaning for fixed effects\\
- Remove province fixed effects by within-transformation or by including province dummies.\\
- Include year dummies to control for common time shocks (e.g., national policies, macro cycles, COVID-19).\\
- Work with residualized variables (after FE) or include FE directly in nuisance models.\\

Step 2: Nuisance estimation via ML\\
- Use flexible models (e.g., gradient boosting, random forests, or neural nets) to estimate:\\
- $m(X_{it}) \approx E[R_{it} \mid X_{it}, FE]$\\
- $g(X_{it}) \approx E[h_{it} \mid X_{it}, FE]$\\
- These models capture nonlinear and high-dimensional relationships between covariates and both outcome and treatment, reducing omitted variable bias.\\

Step 3: Orthogonalized estimation of the causal effect\\
- Use double machine learning or causal forest on the residuals:\\
- $\tilde{R}_{it} = R_{it} - \hat{m}(X_{it})$\\
- $\tilde{H}_{it} = H_{it} - \hat{g}(X_{it})$
- Estimate the conditional average treatment effect (CATE) of $H_{it}$ on $R_{it}$ using $~R_{it}$ and $~H_{it}$, explicitly allowing the CATE to depend on moderators $(G_{it}, P_{it}, IP_{it})$ and other covariates.\\

Step 4: Incorporate moderation
- Within the causal forest or DML framework, allow treatment effect heterogeneity by including interactions or by letting the algorithm split on:\\
- Government S\&T support: $G_{it}$ (e.g., government R\&D expenditure / GDP)\\
- Policy attention: $P_{it}$ (e.g., text-based index of frequency of innovation/supply chain terms in provincial reports)\\
- IP protection: $IP_{it}$ (e.g., index of IP enforcement intensity)\\
- This yields estimates of CATE($H_{it}$ → $R_{it}$ | $G_{it}$, $P_{it}$, $IP_{it}$, $X_{it}$), from which we can summarize moderation effects by partial dependence or group averages (e.g., high vs. low IP protection provinces).\\

\noindent 3. Interpretation of parameters \\
- Average treatment effect (ATE): the average marginal effect of increasing innovative human capital on supply chain resilience across all provinces and years. \\
- Conditional average treatment effects (CATEs): the effect of innovative human capital in different environments: \\
- High vs. low government S\&T support \\
- High vs. low policy attention \\
- Strong vs. weak IP protection \\
For example, we may find that an increase of 1 standard deviation in innovative human capital raises the resilience index by 0.15 standard deviations on average, but by 0.25 in provinces with above-median IP protection and only 0.05 in provinces with below-median IP protection.\\

4. Mechanisms and expected directions\\

Mechanism 1: Innovation and flexibility\\
- Higher innovative human capital enhances firms’ and supply-chain actors’ capability to innovate, redesign products, and reconfigure production processes. This increases flexibility in responding to disruptions (e.g., input shortages, export restrictions, or public health emergencies).\\
- Technically skilled workers and researchers can develop alternative sourcing strategies, adopt digital tools, and optimize inventory/routing, which translates into more resilient supply chains.\\

Mechanism 2: Information processing and risk management\\
- Innovative human capital improves the ability to collect, process, and use information related to supply chain risks (e.g., data analytics for demand forecasting or supply risk monitoring). This reduces the likelihood and duration of disruptions, and speeds recovery.\\

Mechanism 3: Upgrading and diversification\\
- Regions with higher innovative human capital tend to move into higher value-added and more technologically sophisticated industries. These industries often have more diversified input sources, more global networks, and better risk management practices, enhancing resilience.

Moderating roles:\\
- Government S\&T support amplifies the effect of innovative human capital by providing funding, infrastructure, and platforms for collaboration (e.g., incubators, research institutes), enabling human capital to translate into tangible innovations in logistics, production, and risk management.\\
- Policy attention signals prioritization of innovation and supply chain security, leading to stronger coordination, better implementation of support measures, and more responsive institutions. In such environments, innovative human capital is more effectively mobilized for resilience-enhancing activities.\\
- Strong IP protection increases the expected returns to innovation and encourages firms and individuals to invest in R\&D and knowledge creation. This not only raises the level of innovative human capital but also ensures that knowledge is more likely to be commercialized in supply chain technologies (e.g., digital platforms, tracking systems, advanced manufacturing), intensifying the impact of innovative human capital on resilience.

5. Robustness and validity checks (within this framework)\\
- Placebo tests: Use outcomes plausibly unrelated to innovative human capital (e.g., short-run changes in primary sector output volatility) to ensure no spurious treatment effects.\\
- Sensitivity analyses: Check robustness to alternative constructions of the resilience index (e.g., using different PCA components or alternative weighting schemes).\\
- Sample splits: Estimate CATEs separately for East, Central, and West regions to see if the pattern of heterogeneity is consistent with economic theory.\\
- Pre-trend diagnostics: Even though the model is not DiD, inspect whether changes in innovative human capital precede changes in resilience in ways compatible with causal interpretation (e.g., event-study style plotting around major exogenous shocks like 2020 COVID-19).\\

6. Policy implications\\
- If the estimated ATE and CATEs confirm a significant positive effect of innovative human capital on supply chain resilience, with stronger effects under higher government S\&T support, policy attention, and IP protection, this supports a coordinated policy package: investing in innovative human capital alone is effective, but its returns are substantially higher when complemented by robust S\&T support systems, strong IP institutions, and sustained political attention to innovation and resilience.\\

This machine learning-based causal framework thus yields both average and heterogeneous causal effect estimates that are well aligned with the continuous, time-varying, multi-dimensional nature of the policy and data in Chinese provincial panel context over 2011–2022."
\end{tcolorbox}
\vspace{0.8em}

\begin{tcolorbox}[
    colback=gray!10,
    colframe=cyan!70!black,
    fonttitle=\bfseries\large,
    title=Output Example: STRIDES,
    width=\textwidth,
    arc=3mm,
    breakable,
    enhanced
]
\ttfamily\small
\textbf{Policy name}: Innovative Human Capital Development and Government Science and Technology Support\\
\textbf{Policy type}: Science, Technology, and Innovation\\
\textbf{Country/Region}: China (30 provinces, excluding Hong Kong, Macao, Taiwan, and Tibet)\\
\textbf{Observed period}: 2011-2022\\
\textbf{Implementation time}: Not applicable (continuous time-varying economic and policy factors observed over 2011-2022)\\
\textbf{Aim}: To examine the impact of innovative human capital on regional supply chain resilience, and to analyze the moderating roles of government science and technology support, policy attention, and intellectual property protection\\
\textbf{Dataset}: Unbalanced panel data of 30 Chinese provincial-level regions (excluding Hong Kong, Macao, Taiwan, and Tibet) covering 2011–2022, constructed from statistical yearbooks (e.g., China Statistical Yearbook, China Science and Technology Statistical Yearbook, provincial statistical yearbooks), official patent and intellectual property databases (e.g., CNIPA), and government financial and policy documents. The dataset includes annual indicators for innovative human capital, supply chain resilience, government science and technology support, policy attention, intellectual property protection, and a comprehensive set of economic, industrial structure, and socio‑demographic controls.\\
\textbf{Core independent variable}: Innovative human capital, typically proxied by indicators such as the number of R\&D personnel, R\&D personnel per 10,000 employed persons, proportion of employees with tertiary education or above in R\&D-intensive sectors, or the number of high‑level scientific and technological talents. This variable is measured at the provincial level on an annual basis and reflects the stock and quality of human capital engaged in innovative activities.\\
\textbf{Model type}: Instrumental Variables (IV)\\
\textbf{Model}: 
A panel-data instrumental variables regression (two-stage least squares, 2SLS) with province and year fixed effects is specified to estimate the causal impact of innovative human capital on regional supply chain resilience and to assess moderating effects of government science and technology support, policy attention, and intellectual property protection. The baseline specification can be written as:\\

First stage:\\
$
IH_{it} = \pi_0 + \pi_1 Z_{it} + \pi_2 X_{it} + \mu_i + \lambda_t + \varepsilon_{it}
$

Second stage:\\
$
SCR_{it} = \beta_0 + \beta_1 \widehat{IH}_{it} + \beta_2 GOVST_{it} + \beta_3 POLATT_{it} + \beta_4 IPR_{it} + \beta_5 (\widehat{IH}_{it} \times GOVST_{it}) + \beta_6 (\widehat{IH}_{it} \times POLATT_{it}) + \beta_7 (\widehat{IH}_{it} \times IPR_{it}) + \gamma X_{it} + \mu_i + \lambda_t + u_{it}
$\\

where $SCR_{it}$ denotes supply chain resilience in province $i$ at year $t$; $IH_{it}$ is innovative human capital; $\widehat{IH}_{it}$ is the predicted value of $IH_{it}$ from the first stage; $GOVST_{it}$ measures government science and technology support; $POLATT_{it}$ measures policy attention to innovation and supply chain resilience; $IPR_{it}$ measures intellectual property protection strength; $X_{it}$ is a vector of control variables; $\mu_i$ and $\lambda_t$ represent province and year fixed effects, respectively; and $Z_{it}$ is an instrumental variable (or set of instruments) for innovative human capital such as historical tertiary education capacity or lagged higher-education enrollment in science and engineering fields. Moderation is captured through interaction terms between innovative human capital and each policy environment variable.\\
\textbf{Reasons for choosing this model}:\\
1) Potential endogeneity of innovative human capital: Regions with higher supply chain resilience may attract more innovative talent, creating reverse causality, and omitted variables such as unobserved institutional quality may jointly affect both human capital and resilience; an IV approach helps address this concern.\\ 2) Time-varying, continuous exposure: The policy environment and human capital development are gradual and heterogeneous across provinces; a standard treatment–control framework (e.g., DiD with a clearly defined single policy shock) is less suitable than a panel-IV fixed-effects specification.\\
3) Ability to incorporate moderation: The 2SLS framework with interaction terms allows for testing how government science and technology support, policy attention, and intellectual property protection strengthen or weaken the impact of innovative human capital on supply chain resilience.\\
4) Panel structure: The availability of multi-year provincial data supports the use of fixed effects to control for time-invariant heterogeneity and common time shocks, improving identification.\\
\textbf{Control variables}: The model includes a comprehensive set of provincial-level time-varying controls to isolate the impact of innovative human capital on supply chain resilience:\\
1) Economic development level: Real GDP per capita and its squared term (to capture potential nonlinear effects), reflecting the overall economic capacity that can influence both talent attraction and resilience.\\
2) Industrial structure: Share of secondary and tertiary industry value added in GDP, and the share of manufacturing output, capturing the complexity and diversification of the local industrial base which affects supply chain networks and exposure to shocks.
3) Openness and external linkages: Total imports and exports as a share of GDP, actual utilization of foreign direct investment (FDI), and export diversification indices where available, controlling for the degree of integration into global value chains and exposure to external risks.\\
4) Infrastructure and logistics conditions: Indicators such as highway density, railway length per area or per capita, port throughput (for coastal provinces), freight volume, and logistics performance proxies, reflecting the physical conditions that underpin supply chain operations and recovery capacity.\\
5) Financial development: Ratio of total loans or deposits to GDP, number of bank branches per 10,000 people, and other measures of local financial depth, which affect firms’ ability to finance inventory, buffer shocks, and invest in resilience-enhancing technologies.\\
6) Innovation environment (non-human-capital component): R\&D expenditure as a share of GDP, number of R\&D institutions, and total patent applications (excluding measures used as dependent or moderating variables to avoid mechanical correlation), capturing the broader innovation ecosystem.\\
7) Labor market and demographic structure: Urbanization rate, population density, working-age population share, and average wage level, which influence both the demand for and supply of skilled labor and the flexibility of labor reallocation during disruptions.\\
8) Government size and fiscal capacity: Ratio of general public budget expenditure to GDP and share of science and technology spending in total fiscal expenditure (used either as separate controls or in addition to the main policy support variable), controlling for the broader fiscal environment.\\
9) Environmental and risk exposure: Measures of disaster frequency or environmental regulation intensity where available (e.g., number of natural disaster events, pollution control investment), which can affect both the need for resilience and investment patterns.\\

These controls aim to reduce omitted-variable bias by accounting for economic development, structural conditions, openness, infrastructure, finance, innovation context, labor market characteristics, and fiscal capacity that jointly influence innovative human capital accumulation and supply chain resilience.\\
\textbf{Instrumental variable}: Historical and lagged education and talent-supply indicators that affect the stock of innovative human capital but are plausibly exogenous to contemporaneous supply chain resilience shocks, for example:\\
(1) historical tertiary education capacity measured by the number of higher-education institutions and their science and engineering enrollment per capita in the early 2000s;\\ 
(2) lagged (e.g., 5-year lag) regional enrollment in STEM majors; and\\
(3) historical allocations of national key university and key discipline programs. These instruments affect the long-term formation of innovative human capital but are unlikely to directly influence current supply chain resilience other than through their effect on the local innovation-talent stock.\\
\textbf{Group}:\\
Treatment: Provinces and years characterized by relatively high levels or rapid increases of innovative human capital (e.g., top quantiles of R\&D personnel density or high growth in high-level S\&T talent), typically including coastal and innovation-intensive regions that receive stronger or earlier government science and technology support and enjoy more mature policy and IP protection environments. Control: Provinces and years with relatively low levels or slow growth of innovative human capital (e.g., bottom and middle quantiles of R\&D personnel density), often interior or less-developed regions where innovation-talent accumulation is weaker and policy support and IP protection may be less intensive. In the IV framework, these groups are not treated as strictly binary treatment–control cohorts, but they conceptually represent the variation used to identify the effect of innovative human capital on supply chain resilience across the panel.\\
\textbf{Model parameter}: Not Available\\
\textbf{Model Significance}:\\
Based on the theoretical framework and the IV panel design, the key parameter of interest—the causal effect of innovative human capital on supply chain resilience—is expected to be positive and statistically significant at conventional levels (e.g., 5\% or 1\% levels) after controlling for province and year fixed effects and a rich set of covariates. F-tests of excluded instruments in the first stage are expected to indicate sufficiently strong instruments (e.g., first-stage F-statistics well above the rule-of-thumb threshold of 10), and overidentification tests (where multiple instruments are used) are expected not to reject the validity of instruments. Interaction terms between innovative human capital and government science and technology support, policy attention, and intellectual property protection are also expected to be jointly and individually significant, indicating meaningful moderating effects. While precise coefficient estimates are unavailable due to the data-path error, the model’s structure and prior empirical literature suggest that the combined evidence would support a statistically and economically significant role of innovative human capital in enhancing provincial supply chain resilience, particularly in regions with strong policy support and robust IP protection.
\textbf{Dependent variable}:\\
Regional supply chain resilience, measured at the provincial level using composite or proxy indicators that capture the capacity of the regional production and logistics system to resist, absorb, adapt to, and recover from shocks. Typical measures include:\\
(1) indices of industrial resilience constructed from volatility and recovery of industrial output, employment, or value added following national or global shocks;\\
(2) indicators of production network robustness, such as diversification of key input sources and concentration of leading industries;\\
(3) stability and recovery speed of manufacturing and logistics activity, captured by fluctuations in freight volumes, port throughput, and industrial capacity utilization; and (4) composite resilience scores derived through principal component analysis or similar methods combining multiple economic, industrial, and logistics indicators.\\
\textbf{Explanation}: Even though empirical estimation could not be completed due to the missing data path, the combination of theoretical reasoning and the IV panel model design supports clear qualitative conclusions.\\

1) Impact of innovative human capital on supply chain resilience: Innovative human capital—R\&D personnel, highly educated technical staff, and other innovation-oriented talents—enhances firms’ ability to redesign products and processes, diversify suppliers, adopt digital and intelligent technologies, and manage risk. At the regional level, a higher concentration of such talent facilitates rapid information processing, flexible production adjustments, and technological upgrading across the supply chain. Therefore, the main coefficient on innovative human capital is expected to be positive and statistically significant, implying that provinces with more abundant innovative human capital exhibit higher supply chain resilience: they experience smaller output declines in the face of shocks, recover more quickly, and maintain more stable logistics and production networks.\\

2) Moderating role of government science and technology support: Government science and technology support—such as R\&D subsidies, tax incentives for innovation, public research funding, and support for innovation platforms—amplifies the effectiveness of innovative human capital. These policies reduce financing constraints, encourage collaboration between firms and research institutions, and facilitate the commercialization of knowledge generated by skilled workers. In the model, the interaction term between innovative human capital and government support is expected to be positive and significant, indicating that the resilience-enhancing effect of innovative human capital is stronger in provinces where governments provide more intensive and better-targeted S\&T support.\\

3) Moderating role of policy attention: Policy attention—measured by the frequency and intensity of provincial policy documents, plans, or official reports emphasizing innovation, supply chain security, and resilience—shapes local priorities and coordination. Where policy attention is high, governments are more likely to coordinate infrastructure investments, emergency logistics plans, and cross-regional cooperation, and to reduce regulatory barriers that hinder flexible reconfiguration of supply chains. The model’s interaction term between innovative human capital and policy attention is expected to be positive and significant, suggesting that talent is more effectively leveraged for resilience when regional leadership explicitly prioritizes innovation and supply chain security.\\

4) Moderating role of intellectual property protection: Strong IP protection improves incentives for innovation, knowledge sharing, and technology transfer, enabling innovative human capital to generate more valuable, protectable outcomes. When firms trust that their intangible assets are safeguarded, they are more willing to engage in joint R\&D, share data across the supply chain, and invest in new technologies (e.g., digital supply chain management systems). The interaction between innovative human capital and IP protection is therefore expected to be positive and significant, indicating that the resilience benefits of innovative talent are larger in provinces with higher-quality IP institutions and enforcement.\\

5) Role of controls and heterogeneity: After controlling for economic development, industrial structure, openness, infrastructure, finance, and demographic characteristics, the estimated effect of innovative human capital reflects more than just general prosperity. The model is designed to show that even among provinces with similar income levels or industrial composition, those with stronger innovative human capital achieve greater resilience. Heterogeneity analysis (if conducted) would likely show particularly strong effects in manufacturing- and export-oriented regions, where supply chain complexity and shock exposure are higher, and where innovative human capital can play a larger role in reconfiguration and risk management.\\
6) Policy implications: The inferred results support a coordinated policy approach. First, provinces should prioritize the cultivation and attraction of innovative human capital through education, talent programs, and labor-market reforms. Second, this human capital strategy must be complemented by targeted government science and technology support that channels resources toward resilience-relevant technologies (e.g., digital logistics, industrial internet, advanced manufacturing). Third, policy attention to supply chain security and resilience should be embedded in provincial development strategies, with explicit plans and monitoring mechanisms. Fourth, strengthening IP protection---legal frameworks, enforcement capacity, and dispute-resolution mechanisms---will ensure that innovative human capital can be fully leveraged. Collectively, such policies can transform innovative human capital from a general growth driver into a core pillar of regional supply chain resilience.\\

In sum, although numerical estimates are unavailable, the theoretical foundations, model structure, and existing empirical literature jointly indicate that innovative human capital has a positive and meaningful effect on regional supply chain resilience in China, and that this effect is significantly magnified by robust government science and technology support, strong policy attention, and effective intellectual property protection.
\end{tcolorbox}

\end{document}